%% file: OoDAnalyzer-TVCG.tex
\documentclass[10pt,journal,compsoc]{IEEEtran}
\ifCLASSOPTIONcompsoc
  \usepackage[nocompress]{cite}
\else
  \usepackage{cite}
\fi

\ifCLASSINFOpdf
  \usepackage[pdftex]{graphicx}
\else
  \usepackage[dvips]{graphicx}
  \DeclareGraphicsExtensions{.eps}
\fi

\usepackage{mathptmx}
\usepackage[pdftex]{graphicx}
\usepackage{times}
\usepackage{paralist}
\usepackage{amsmath}
\usepackage{url}
\usepackage{cite}
\usepackage{array}

\usepackage{bm}
\usepackage{amssymb}
\usepackage{verbatim}
\usepackage{multirow}
\usepackage{makecell}
\usepackage{ulem} 
\usepackage{tikz}
\usepackage{color,microtype,hyphenat,balance}
\usepackage{graphicx,subfigure}
\usepackage{booktabs}
\usepackage{enumitem}
\usepackage{amsthm}
\usepackage{textcomp}
\usepackage{gensymb}
\usepackage{algorithmicx}
\usepackage{overpic}
\usepackage{contour}
\usepackage{ragged2e}
\usepackage{rotating}

\usepackage{xcolor}
\usepackage[english]{babel}
\usepackage{wrapfig}
\usepackage{ulem}
\usepackage[hypcap=false]{caption} 

\usepackage[ruled,vlined]{algorithm2e}
\DeclareMathAlphabet{\mathcal}{OMS}{cmsy}{b}{n}
\DeclareMathAlphabet{\mathcal}{OMS}{cmsy}{m}{n}

\newtheorem{theorem}{Theorem}

\def \etal {{\emph{et al}.\thinspace}}

\def \eg {{\emph{e.g}.\thinspace}}
\def \ie {{\emph{i.e}.\thinspace}}

\newcommand{\myparagraph}[1]{\vspace{1mm}\noindent\textbf{#1}}

\def \mx {\mathbf{x}}
\def \my {\mathbf{y}}

\def \md {\mathrm{deg}}

\newcommand{\doc}[1]{\textcolor{black}{#1}}
\newcommand{\shixia}[1]{\textcolor{black}{#1}}
\newcommand{\xia}[1]{\textcolor{black}{#1}}
\newcommand{\yafeng}[1]{\textcolor{black}{#1}}
\newcommand{\changjian}[1]{\textcolor{black}{#1}}
\newcommand{\yuanjun}[1]{\textcolor{black}{#1}}
\newcommand{\grammarly}[1]{\textcolor{black}{#1}}

\newcommand{\gramsecondrev}[1]{\textcolor{black}{#1}}
\newcommand{\docsecondrev}[1]{\textcolor{black}{#1}}
\newcommand{\cjsecondrev}[1]{\textcolor{black}{#1}}

\usepackage[bookmarks,backref=true,linkcolor=black]{hyperref} 

\ifCLASSOPTIONcompsoc
\else
\fi

\ifCLASSINFOpdf
\else
\fi

\begin{document}
%
\title{OoDAnalyzer: Interactive Analysis of Out-of-Distribution Samples}
\author{
Changjian Chen*, Jun Yuan*, Yafeng Lu, Yang Liu, Hang Su, Songtao Yuan, Shixia Liu
\IEEEcompsocitemizethanks{
\IEEEcompsocthanksitem C.~Chen, J.~ Yuan, H.~Su, and S.~Liu are with School of Software, Dept.~of Comp.~Sci.~\& Tech., BNRist, Tsinghua University. ${}^*$Equal contribution, S.~ Liu is the corresponding author.
\IEEEcompsocthanksitem Y.~Lu is with Bloomberg L.P; Y.~Liu is with Microsoft Research Asia.
\IEEEcompsocthanksitem S.~Yuan is with the First Affiliated Hospital of Nanjing Medical University.

}
\thanks{}}
%


\teaser{
\setcounter{figure}{0}
\centering
\vspace{-14pt}
\begin{overpic}[width=0.9\linewidth]{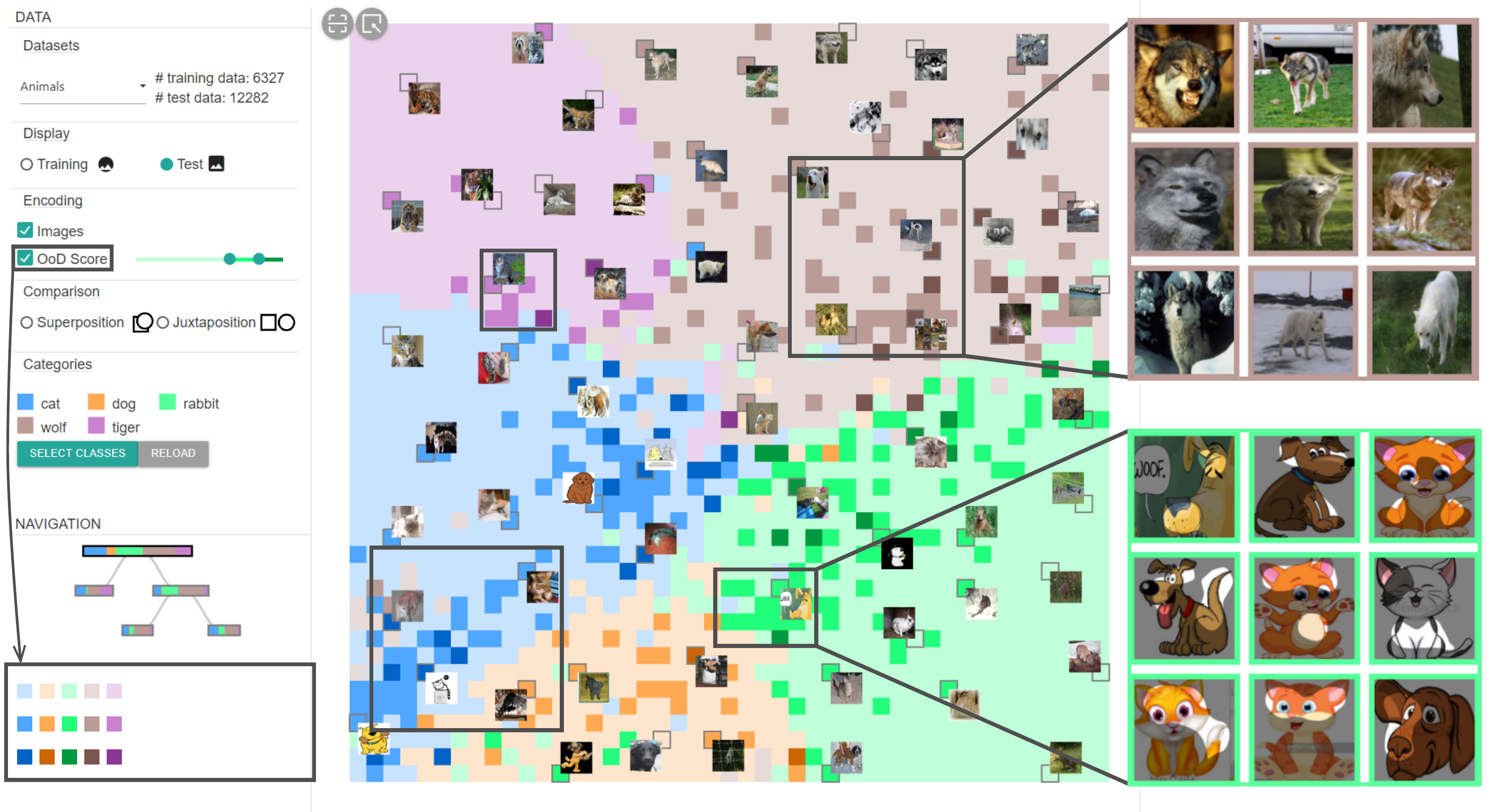}
\put(25.5,15){\contour[32]{white}{\textcolor{black}{A}}}
\put(48.5,13.5){\contour[32]{white}{\textcolor{black}{B}}}
\put(62,41){\contour[32]{white}{\textcolor{black}{C}}}
\put(33,33){\contour[32]{white}{\textcolor{black}{D}}}
\put(20,43){\contour[32]{white}{\textcolor{black}{E}}}
\put(20,37){\contour[32]{white}{\textcolor{black}{F}}}
\put(20,32){\contour[32]{white}{\textcolor{black}{G}}}
\put(20,26){\contour[32]{white}{\textcolor{black}{H}}}
\put(20,12){\contour[32]{white}{\textcolor{black}{I}}}
\put(8.7,7.8){\scriptsize Low OoD score}
\put(8.7,5.5){\scriptsize Middle OoD score}
\put(8.7,3){\scriptsize High OoD score}
\put(7,-1){(a)}
\put(47,-1){(b)}
\put(87,26.5){(c)}
\put(87,-1){(d)}
\end{overpic}
\vspace{2pt}
\captionsetup{width=0.9\linewidth}
\caption{
A visual analysis approach for understanding Out-of-Distribution (OoD) samples caused by \shixia{the distribution difference} between training and test data: 
(a) controls and visualization aids; 
(b) a grid-based visualization to illustrate OoD \changjian{samples} in context. The color of the grid \changjian{cell and border} encodes the category, 
and the sequential color scheme represents the \changjian{OoD scores of samples}; \shixia{(c) image and (d) saliency map representations of the selected region}.}
\label{fig:teaser}
\vspace{-5mm}
}

\input{abstract}

\maketitle
{
\fontsize{10}{10} 
\input{introduction}

\input{related}

\input{motivation-design-system.tex}

\input{data-modeling.tex}

\input{visualization.tex}

\input{application.tex}

\input{discussion.tex}





}

\IEEEdisplaynotcompsoctitleabstractindextext

%
\IEEEpeerreviewmaketitle


%

\ifCLASSOPTIONcompsoc
  \section*{Acknowledgments}
\else
  \section*{Acknowledgment}
\fi

C. Chen, J. Yuan, and S. Liu are supported by the National Key R\&D Program of China (No. 2018YFB1004300) and the National Natural Science Foundation of China (No.s 61936002, 61761136020, 61672308).
H. Su is supported by the National Key  R\&D Program of China (No. 2017YFA0700904), NSFC Projects (Nos. 61620106010, 61621136008), and the JP Morgan Faculty Research Program.
\cjsecondrev{The authors would like to thank Shen Wang for providing \xia{the} source code of the UNet++ model and Keren Xie for his contribution to REA data analysis.}

\ifCLASSOPTIONcaptionsoff
  \newpage
\fi



%
\bibliographystyle{IEEEtran}
\bibliography{reference}

\begin{IEEEbiography}[{\includegraphics[width=1in,height=1.25in,clip,keepaspectratio]{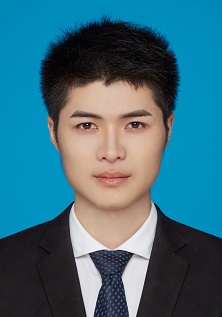}}]
{Changjian Chen} is now a Ph.D. student at Tsinghua University. His research interests are in interactive machine learning. He received a B.S. degree from University of Science and Technology of China.
\end{IEEEbiography}

\begin{IEEEbiography}[{\includegraphics[width=1in,height=1.25in,clip,keepaspectratio]{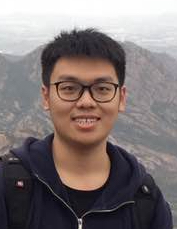}}]{Jun Yuan}
is currently a Ph.D. student at Tsinghua University. His research interests are in explainable artificial intelligence. He received a B.S. degree from Tsinghua University. 
\end{IEEEbiography}

\begin{IEEEbiography}[{\includegraphics[width=1in,height=1.25in,clip,keepaspectratio]{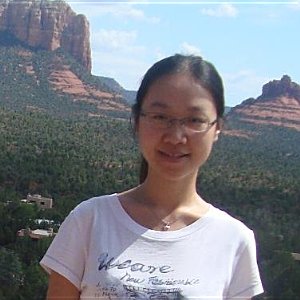}}]
{Yafeng Lu} is a software engineer at Bloomberg L.P.  Prior to joining Bloomberg, she was a Postdoctoral Research Associate in the School of Computing, Informatics and Decision Systems Engineering at Arizona State
University. Her research interests are in predictive visual analytics.
\end{IEEEbiography}

\begin{IEEEbiography}[{\includegraphics[width=1in,height=1.25in,clip,keepaspectratio]{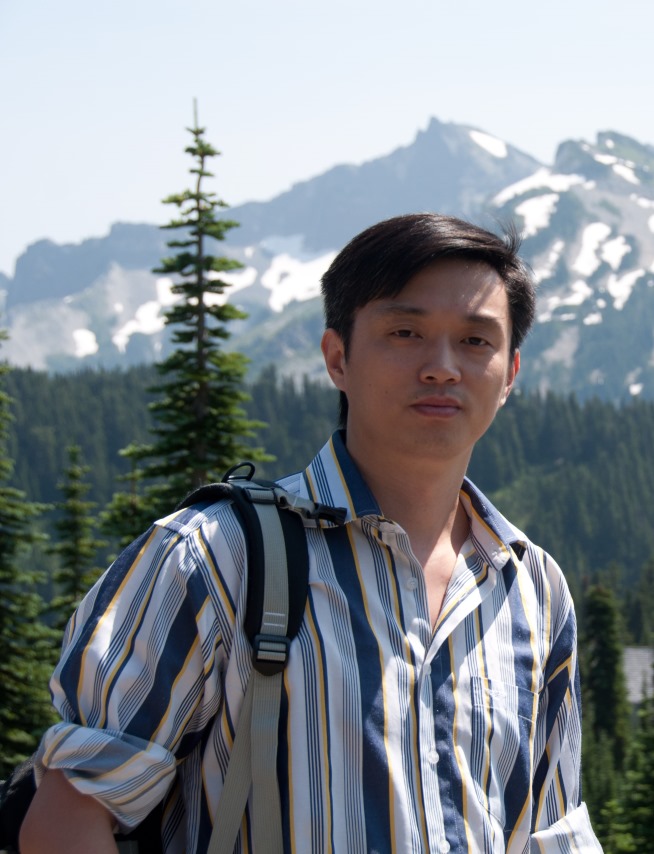}}]
{Yang Liu}
is a principal researcher at Microsoft Research Asia. He received his Ph.D. degree from The University of Hong Kong, Master and Bachelor degrees from University of Science and Technology of China. His recent research focuses on geometric computation and learning-based geometry processing and generation. He is an associate editor of IEEE Trans. Vis. Comput. Graph.
\end{IEEEbiography}

\begin{IEEEbiography}[{\includegraphics[width=1in,height=1.25in,clip,keepaspectratio]{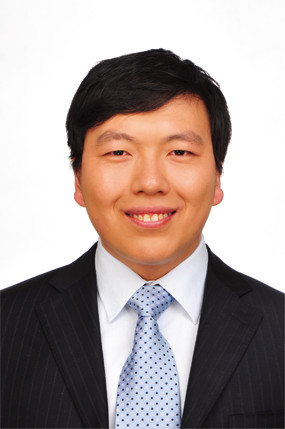}}]
{Hang Su} is an assistant professor at Tsinghua University. He received his B.S, M.S., and Ph.D. Degrees from Shanghai Jiaotong University. His research interests lie in the development of computer vision and machine learning algorithms for solving scientific and engineering problems arising from artificial learning, reasoning, and decision-making. His current work involves the foundations of interpretable machine learning and the applications of image/video analysis. He has served as senior PC members in the dominant international conferences.
\end{IEEEbiography}

\begin{IEEEbiography}
[{\includegraphics[width=1in,height=1.25in,clip,keepaspectratio]{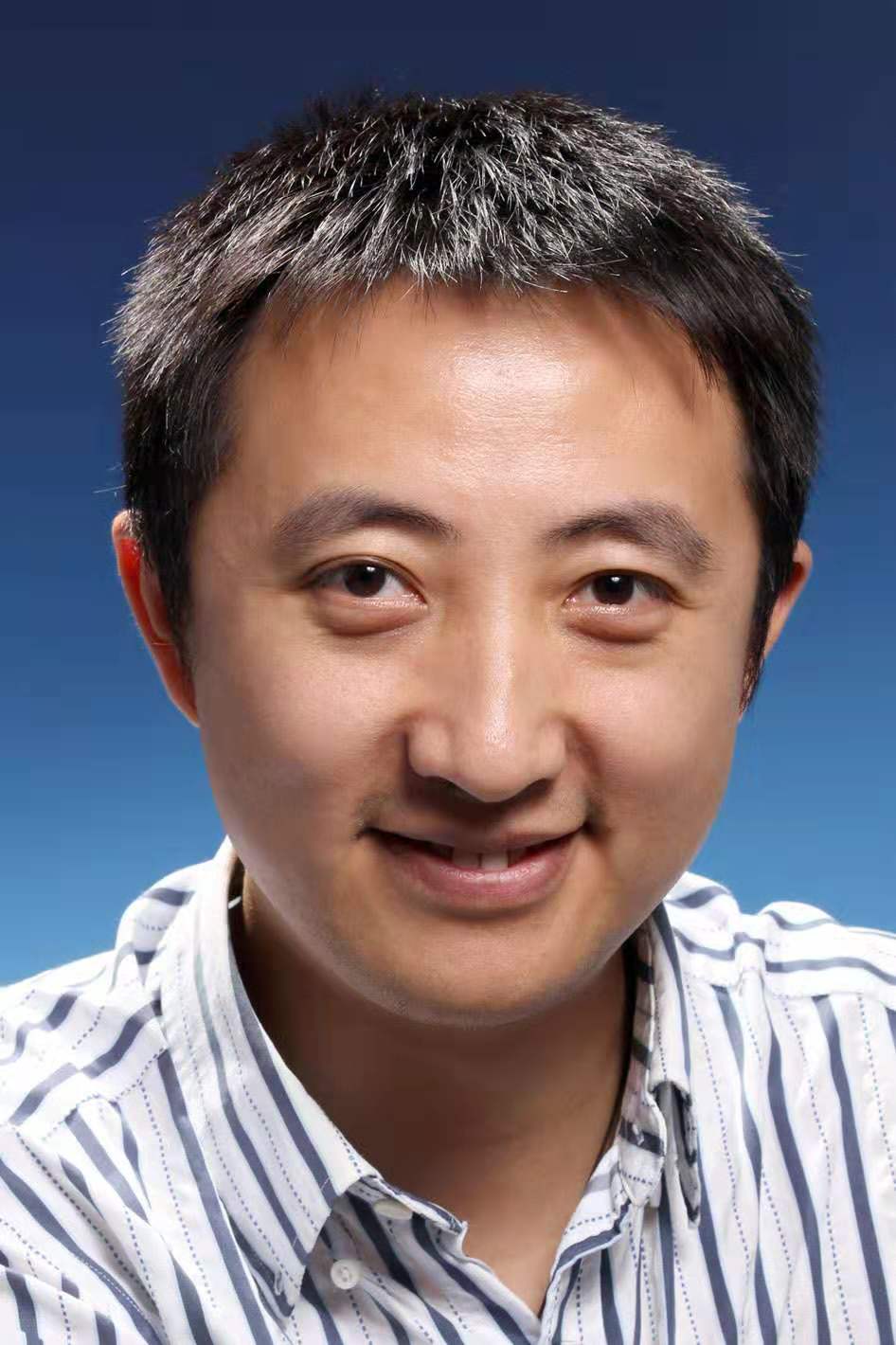}}]{Songtao Yuan}, MD. Retina specialist, Ophthalmology Department, The first affiliated hospital of Nanjing medical university, Nanjing, Jiangsu Province, China. Major research interests are degenerative diseases of retina and stem cells regeneration. 
\end{IEEEbiography}

\begin{IEEEbiography}
[{\includegraphics[width=1in,height=1.25in,clip,keepaspectratio]{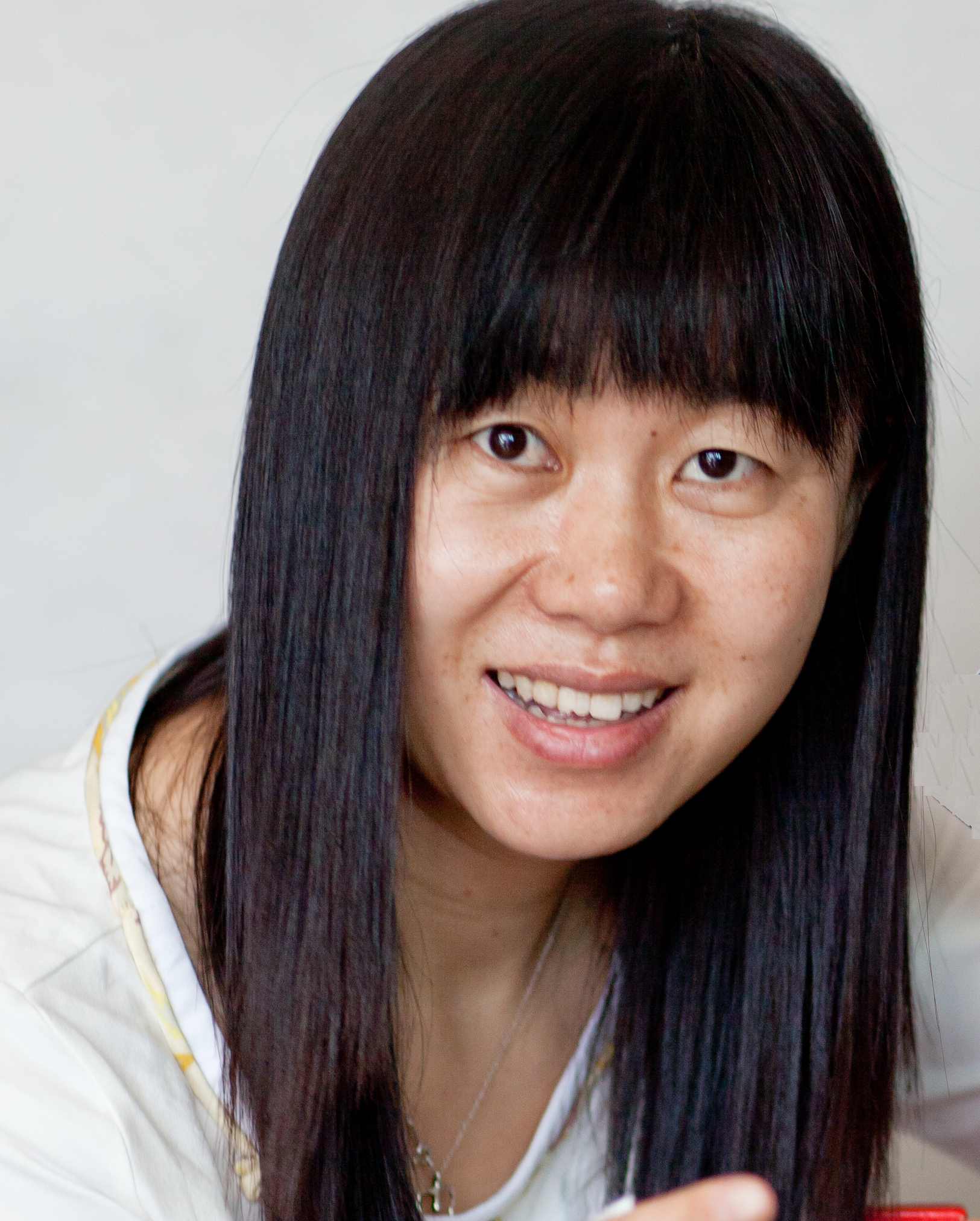}}]{Shixia Liu}
is an associate professor at Tsinghua University. Her research interests include visual text analytics, visual social analytics, interactive machine learning, and text mining. She worked as a research staff member at IBM China Research Lab and a lead researcher at Microsoft Research Asia.
She received a B.S. and M.S. from Harbin Institute of Technology, a Ph.D. from Tsinghua University.
She is an associate editor-in-chief of IEEE Trans. Vis. Comput. Graph.
\end{IEEEbiography}

\end{document}


\maketitle
\input{appendix.tex}

\bibliographystyle{IEEEtran}
\bibliography{appendix-ref}

%% file: abstract.tex
\IEEEcompsoctitleabstractindextext{
\justify
\begin{abstract}
One 
\yafeng{major cause}
of performance degradation in predictive models is that the test samples are not well covered by the training data.
Such not well-represented samples are called OoD samples.
In this paper,
we propose OoDAnalyzer, a visual analysis approach for interactively identifying OoD samples and 
\shixia{explaining them in context.}
Our approach integrates an ensemble OoD detection method and a grid-based visualization.
\yafeng{The detection method is improved from deep ensembles by combining more features with algorithms in the same family. }
To better analyze and understand the OoD \changjian{samples} in context, we have developed a novel $k$NN-based grid layout algorithm motivated by Hall's theorem.
The algorithm approximates the optimal layout and has $O(kN^2)$ time complexity, faster than \shixia{the grid layout algorithm with overall best performance but $O(N^3)$ time complexity.}
Quantitative evaluation and case studies were performed on several datasets to demonstrate the effectiveness \shixia{and usefulness} of OoDAnalyzer.
\end{abstract}
\begin{IEEEkeywords}
Out-of-Distribution detection, grid layout, interactive visualization
\end{IEEEkeywords}
}

%% file: introduction.tex

\section{Introduction}
\maketitle 
In machine learning, especially in supervised learning, 
the existence of Out-of-Distribution (OoD) samples, the test samples that are not well covered by training data, 
is \docsecondrev{a major cause} of performance degradation~\cite{hendrycks2016baseline, louizos2017multiplicative, guo2017calibration}.
The model often performs poorly on these OoD samples~\cite{lee2018simple},
and may even misclassify OoD samples with high confidence.
This high-confidence misclassification \docsecondrev{can lead to serious real-world} incidents,
\docsecondrev{such as} self-driving car crashes.
\looseness=-1

\begin{figure}[t]
  \centering
  \begin{overpic}[width=\linewidth]{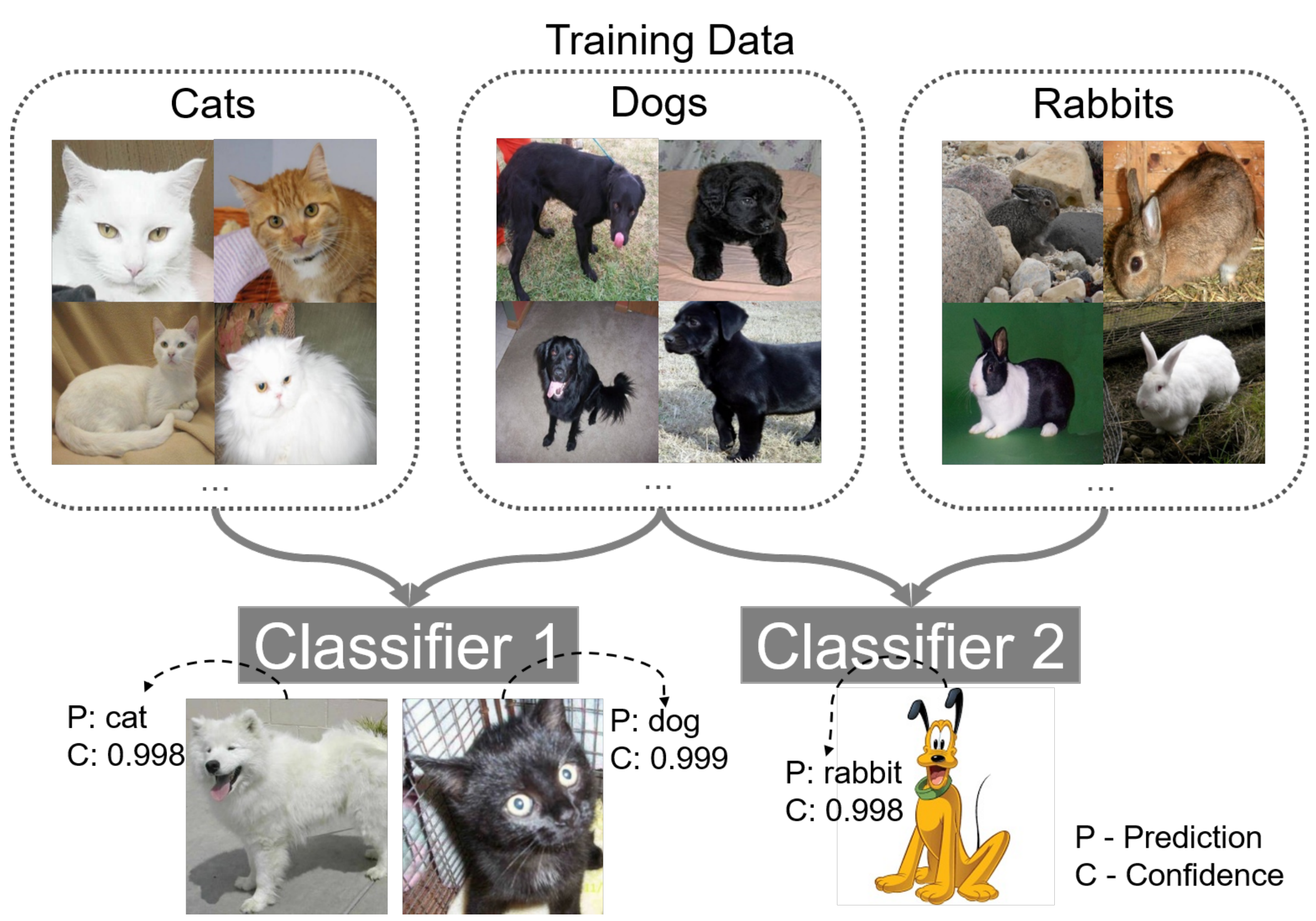}
  \put(28,-3){(a)}
  \put(69,-3){(b)}
  \end{overpic}
    \caption{OoD \changjian{samples} in image classification: (a) a white dog and a black cat are incorrectly predicted with high confidence by a classifier trained on a dataset only consisting of dark-colored dogs and light-colored cats; 
    (b) a more complex case with a cartoon dog.
    }
    \label{fig:bias-dog-cat}
    \vspace{-5mm}    
\end{figure}

To mitigate the problem of OoD samples, 
several detection methods have been developed to identify them in the test dataset~\cite{hendrycks2016baseline, lee2018simple, liang2017enhancing}.
However, challenges still exist \doc{for} retrieving OoD samples and explaining the OoD context.
First, most OoD detection methods either use a single model or a few models of the same hyper-parameters with a variety of high-level features~\cite{lakkaraju2017identifying,lakshminarayanan2017simple}, 
which often fail to cover the whole data distribution and \cjsecondrev{may} discard some key information in the OoD samples. 
Second, by only outputting a confidence value for each sample, 
existing OoD detection methods 
\yafeng{cannot explain the}
underlying reason for the appearance of such OoD samples.
\shixia{Thus,}
it is difficult to systematically address the problem caused by the OoD samples.
For example, a dog-cat image classifier might use color as the key criteria to make predictions when the training dataset \docsecondrev{only has} dark-colored dogs and light-colored cats, 
as illustrated in Fig.~\ref{fig:bias-dog-cat}(a).
Fig.~\ref{fig:bias-dog-cat}(b) shows \shixia{a more complicated example where} a cartoon character could be \yafeng{another} cause to have OoD samples.
\shixia{This is} because
cartoons are always exaggerated \docsecondrev{and} may distort some key information.
In this case, the dog has very big, long ears, which is a key feature of a rabbit.
\changjian{Uncovering} these complicated reasons requires \gramsecondrev{a} more detailed analysis.
\shixia{As a result, an intuitive and self-explanatory exploration environment is needed to comprehensively summarize different reasons and present the findings in an interpretative manner.}

In this work, we propose a visual analysis approach, OoDAnalyzer, which integrates an ensemble OoD detection method with interactive visualizations to help identify OoD samples and explain the OoD context.
Once OoD samples have been identified and \doc{their} root causes \doc{have been determined}, 
labeling and adding samples accordingly to the training dataset can usually improve the model accuracy~\cite{lee2017training,liang2017enhancing}.
Hence OoDAnalyzer mainly focuses on identifying and explaining OoD samples.
OoDAnalyzer starts by recommending some OoD \changjian{samples} using the developed ensemble OoD detection method. 
This method is based on the state-of-the-art OoD detection method (deep ensembles)~\cite{lakshminarayanan2017simple}, but combines more features with more algorithms in a family.
The algorithm family is derived from the same classification algorithm by varying its hyper-parameters.
\yafeng{Our experiments show that this improved detection method}
has higher accuracy than deep ensembles~\cite{lakshminarayanan2017simple} \doc{when it comes to} detecting OoD samples.
With the detected OoD \changjian{samples}, we propose a grid-based visualization to explore them in the context of the training and test datasets.
\yafeng{\doc{A} well-known grid layout algorithm has $O(N^3)$ time complexity and can not support interactive exploration for large datasets. 
We \gramsecondrev{propose} a $k$NN-based approximation and speed it up to $O(kN^2)$ for real-time interactions. }
We use image classification as an example to validate our approach.
Specifically, experiments were conducted to quantitatively evaluate the OoD detection method and \cjsecondrev{the} $k$NN-based grid layout approximation algorithm, and two case studies were conducted on an animal image classification problem and a retinal edema classification problem. 
After augmenting the training data with samples 
complementing the detected OoD samples,
the model accuracy improved greatly.
This demonstrates the effectiveness of our approach. 
A demo of the prototype is available at: \url{https://bit.ly/2W4jvcc}.

In summary, the main contributions of our work are:
\begin{itemize}[leftmargin=*]\setlength\itemsep{0mm}
    \item[--] A visual analysis tool \shixia{that explains OoD \changjian{samples} in context and allows users to analyze them effectively}.
    \item[--] A grid-based visualization with a proposed grid layout acceleration method inspired by Hall's theorem.
    \item[--] A practical improvement of ensemble OoD detection for OoD \changjian{sample} recommendation.
\end{itemize}

%% file: related.tex
\section{Related Work}
\label{sec:related-work}

\subsection{OoD detection}
Attempts at OoD detection \changjian{fall into} two categories: detection of OoD samples with low confidence predictions~\cite{hendrycks2016baseline,guo2017calibration} and detection of OoD samples with high confidence predictions~\cite{lakkaraju2017identifying,lakshminarayanan2017simple}.

Some earlier methods were
intended to identify OoD samples with low confidence scores~\cite{guo2017calibration,hendrycks2016baseline,liang2017enhancing}.
These methods are based on the fact that the prediction distributions are different between normal samples and OoD samples with low confidence scores~\cite{hendrycks2016baseline}.
In particular, Hendrycks \etal~\cite{hendrycks2016baseline} estimated the OoD confidence by the prediction distribution of each sample. 
\changjian{Later research} further improved the detection performance by using temperature scaling~\cite{guo2017calibration}, adding small controlled perturbation~\cite{liang2017enhancing}, and jointly training a generator and a classifier~\cite{lee2017training}. 

The aforementioned methods have achieved some success in detecting OoD samples with low confidence scores that are near the decision boundaries. 
However, in many real-world applications, 
OoD samples with high confidence scores \docsecondrev{can blend in} reliable normal samples, which tends to significantly degrade the model performance~\cite{liang2017enhancing}.
To tackle this issue, 
Lakkaraju \etal~\cite{lakkaraju2017identifying} iteratively utilized an estimated OoD proportion to probabilistically sample a data item from an item group and queried an oracle for its true label.
 After the true label of the item was confirmed, the estimated proportion of this group was updated.
To further improve the accuracy and robustness of OoD detection, 
deep ensembles~\cite{lakshminarayanan2017simple} \doc{was} \docsecondrev{developed} by combining different high-level semantic features (learned by neural networks) with a single classification algorithm.
The discrepancy among different classifiers in deep ensembles is used to measure the OoD degree of a sample.

Deep ensembles provides a practical way to detect OoD samples more effectively. 
Inspired by this method, we have developed an OoD detection method that enumerates the combinations of all features (both low-level and high-level features) with different algorithms in a family.
The extension of the features and algorithms aims \docsecondrev{to reduce} the uncertainty caused by the limited coverage of both the employed features and algorithms in deep ensembles.
In addition, a grid-based visualization has been developed to facilitate the exploration and analysis of the detected OoD \changjian{samples}.

\subsection{Visualization for Outlier Detection}
In the field of visual analytics, 
the \docsecondrev{work most relevant} to ours is visual outlier detection.
It aims \docsecondrev{to identify} samples that are inconsistent with the remainder of that set of data~\cite{barnett1974outliers, liu2014survey}.
Existing efforts can be classified into two categories~\cite{Jiang2018}: sequence-based methods~\cite{liu2018analyzing,lu2016exploring,wang2019a,xu2017vidx,zhao2014fluxflow} and point-based methods~\cite{cao2016targetvue,cao2018zglyph,thom2012spatiotemporal,wilkinson2018visualizing}.
Our work is relevant to point-based methods, so we focus on reviewing this category.

Earlier efforts \doc{developed} visualizations for visually 
presenting and analyzing outliers and anomalies.
For example, Thom~\etal~\cite{thom2012spatiotemporal} displayed anomalous spatial-temporal activities on a map view.
A great deal of later research focused on supporting contextual exploration to interpret the reason for the appearance of
outliers.
Ko~\etal~\cite{ko2014analyzing} enabled users to configure the context in which anomalies were detected.
TargetVue, a visual analysis system detecting anomalous users in Twitter, was proposed by Cao~\etal~\cite{cao2016targetvue}. 
It employed user glyphs to summarize behavior semantics and used a triangle grid layout for user comparison.

Recently, with the growing integration of modeling and user interactions, \shixia{several} visual analysis systems have been developed to support iterative anomaly detection. 
An adaptive anomaly integration mechanism was designed by Cao~\etal~\cite{cao2018voila}, to include human judgment in the detection algorithm for refining the anomalies.
Xie~\etal~\cite{xie2019visual} applied One-Class SVM to detect anomalous call stack trees and used MDS and scatterplot to separate detected trees from normal trees. 
The anomalous call stack trees can be further investigated and verified by users whose verification will then update the anomaly detection model. 
LabelInspect~\cite{liu2019interactive} was developed to emphasize uncertain labels on a constrained t-SNE visualization and allow users to interactively update the classification model and recommend a new set of uncertain labels. 

Similar to previous methods, we also support \doc{context-enriched} exploration, where semantically similar samples are placed close together, and saliency maps are displayed to explain effective regions of the image for the prediction.
In contrast to the aforementioned methods, with the goal of analyzing OoD \changjian{samples}, our approach satisfies two distinct requirements: extensive comparisons between the training and test datasets and explorations loaded with content-level examinations
in large datasets.
To this end, we propose a grid-based visualization that builds upon grid layouts and uses $k$NN approximation to achieve real-time interactions. 
Our visualization keeps semantic similarities between samples and enables \changjian{exploration} of the content.
It also supports dataset comparison via a hybrid of superposition and juxtaposition, as well as detailed exploration in the local context.

%% file: motivation-design-system.tex
\section{Design of OoDAnalyzer}\label{Sec:Design}
\subsection{Requirement Analysis}
\label{sec:requirement}

\begin{figure*}[t]
  \centering
    \includegraphics[width=\linewidth]{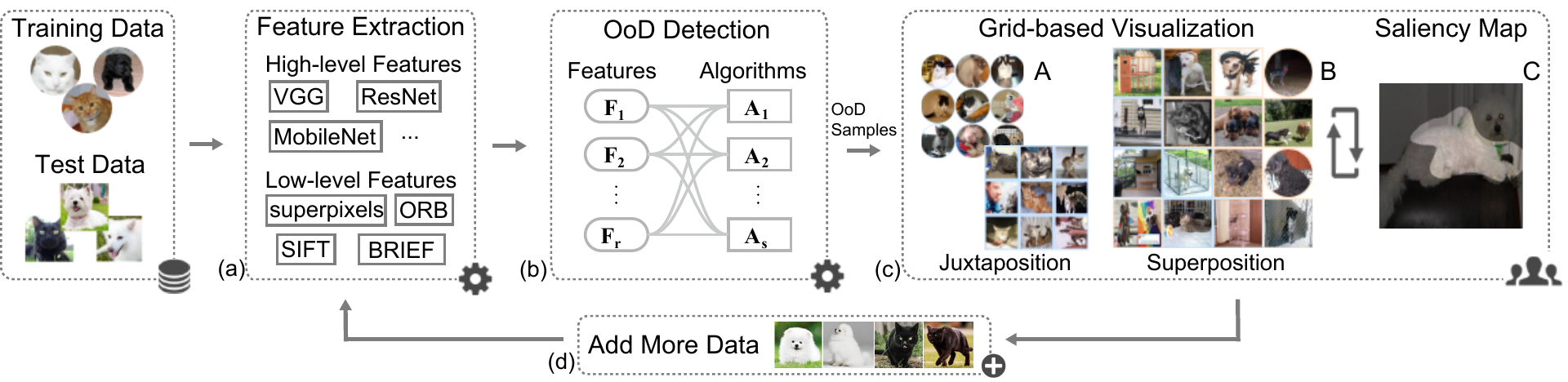}
    \caption{OoDAnalyzer overview. 
     \shixia{(a) Extract high- and low-level features; (b) Identify OoD samples;
    (c) Analyze OoD \changjian{samples} \changjian{and identify the reason why they appear};
    (d) Add more training data \changjian{based on the identified reason} and start a new round of analysis.\looseness=-1
    }
    }
    \label{fig:analysis_pipeline}                                       
    \vspace{-5mm}
\end{figure*}

Recent research has shown
that identifying and illustrating unexpected model failures caused by OoD samples is critical for many practical applications~\cite{lakkaraju2017identifying,lakshminarayanan2017simple,liu2018analyzing, liu2018bridging}.
\yafeng{We \doc{have} also observed this} in our long-term collaboration with machine learning experts and practitioners (\eg, doctors). We have found that the coverage and diversity
of training data is an integral part of supervised learning, 
which determines the upper limits of model performance. 
For instance, Fig.~\ref{fig:bias-dog-cat}(a) describes a real case 
we encountered when collaborating with a machine learning expert on a visual analysis project that seeks to explain the inner workings of deep neural networks \changjian{(DNNs)}. 
As the training dataset only \doc{consisted} of light-colored cats and dark-colored dogs,
the prediction accuracy \doc{degraded} significantly when the test dataset \doc{contained} many white dogs and black cats.
This puzzled us for several weeks. 
After using the t-SNE projection technique and a series of side-by-side comparisons between the training and test datasets, we finally figured out 
that color bias caused the failure.
We also experienced a similar situation when collaborating with two doctors who have employed deep learning models to facilitate retinal edema diagnosis in the past two years.
The aforementioned debugging processes led us to design an interactive tool to help users examine the datasets and their distributions in a failed learning process.  



To identify the primary requirements,
 in our collaboration with two machine learning experts ($M_1$, $M_2$) and two doctors ($D_1$, $D_2$),  
we \docsecondrev{asked them} the following questions. 
 What would you do after you identify a misclassified image? If you find a few similar misclassified images, 
would you like to summarize their \docsecondrev{commonalities}? 
\shixia{What kind of information is helpful in identifying the major cause of misclassified images?}


\changjian{Based on the interviews and the review of the literature, we identified three major requirements.}

\myparagraph{R1 - Examining OoD \changjian{samples} in the context of 
normal samples}. 
In a variety of research and practical projects, our collaborators have found that many OoD \changjian{samples} \doc{were} difficult to confirm 
without comparing \doc{them} to the normal samples in the training dataset. 
They often needed to look at training samples whose \docsecondrev{discriminating} features for the classification task \doc{were} highly similar to those of the OoD \changjian{samples}. 
\yafeng{\doc{T}hey were trying to see the \doc{similarities} and differences between \doc{the} OoD samples and normal samples.}
To this end, the experts \docsecondrev{needed to place} semantically similar images together 
and visually \docsecondrev{represent} each image in the dataset in a compact way, 
\docsecondrev{so} they could easily identify the OoD \changjian{samples} and explore the context in individual cases.  
This requirement is also consistent with the common practice in the current research of identifying dataset bias~\cite{lakkaraju2017identifying,Torralba2011unbaised}.

\myparagraph{R2 - Comparing the datasets and OoD \changjian{samples} in both a global view and local context}.
For a \changjian{poor}-performance predictive model, 
the experts would like to check how OoD \changjian{samples} are distributed over the categories 
and find out which categories \cjsecondrev{may} include real OoD samples by comparing the distributions between the training and test datasets side-by-side.
After identifying the categories of interest, they 
want to narrow down their focus to closely compare the OoD \changjian{samples} to their semantically similar normal samples, which might be found in both the training and test datasets.
Based on \docsecondrev{this} analysis, the experts can then figure out the major reason why these samples are OoD and decide what kind of samples need to be labeled and added to the training dataset to improve the model performance. 
$M_1$ said that for the dog-cat dataset, he carefully examined the training samples that were very close to an incorrectly predicted 
white dog image and found that they were all white cat images.
By further comparing the training data and test data, he noticed that the cats in the training data were exclusively light-colored and thus had a limited color diversity. 
Other experts concurred with $M_1$ on this comparison-based exploration process. \looseness=-1

\myparagraph{R3 - Understanding and analyzing different types of \shixia{OoD \changjian{samples}}}.
Previous studies have shown that OoD \changjian{samples} can be broadly categorized into two types: those with low confidence predictions (known unknowns) and those with high confidence predictions (unknown unknowns)~\cite{lakkaraju2017identifying}. 
Those \docsecondrev{of} the first type are usually near the decision boundaries and have low confidence scores,
while the second type of samples are \docsecondrev{those} which
the model is highly confident in but turn out to be \docsecondrev{misclassified}.
To reduce OoD samples to the maximum extent, all the experts desired to visually explore 
\yafeng{both types of OoD samples and their relationship with other samples.}
This requires employing an OoD detection method to identify both known unknowns and unknown unknowns, and \docsecondrev{presenting} their visual difference in the layout.
\looseness=-1

\subsection{System Overview}

Motivated by these requirements, 
\changjian{we designed OoDAnalyzer (Fig.~\ref{fig:analysis_pipeline}) \shixia{to analyze OoD samples in context}.
Given the training and test data, both high- and low-level features are extracted first (Fig.~\ref{fig:analysis_pipeline}(a)).}
\changjian{These features are then 
fed
into the ensemble OoD detection method to identify OoD samples for further analysis (Fig.~\ref{fig:analysis_pipeline}(b)).}
In particular, 
\changjian{the} detection method learns to detect OoD \changjian{samples} with low and high confidence scores by combining different features with an algorithm family (\textbf{R3}). 
\looseness=-1
However, without a clear understanding of the underlying cause of having these OoD samples, 
it is hard to take any action to improve the model performance.
\looseness=-1

To tackle this issue, 
\changjian{the grid-based visualization was developed to facilitate the exploration and analysis of OoD \changjian{samples} in context (Fig.~\ref{fig:analysis_pipeline}(c)).}
\yafeng{We have considered t-SNE as an alternative solution. However,}
it does not support content-level (\eg, the image itself) exploration due to the overlap among samples. 
Since content-level exploration is important \docsecondrev{for understanding} the OoD context, 
\yafeng{our}
grid-based visualization 
maps the scatter plot generated by t-SNE into a grid layout.
Thus, we can both preserve 
the similarities between \changjian{images} 
and allocate a minimum \doc{of} space to show the image content. 
\shixia{In our implementation, each image is represented by the high-level features extracted \changjian{by} DNNs~\cite{liu2017towards}, which corresponds to a point in the high dimensional feature space.  
The Euclidean distance \doc{between} two points is used to measure the similarity between images. 
Previous studies have shown that features extracted by DNNs can well capture semantic information~\cite{donahue2014decaf, Wang2019Visual}. As a result, t-SNE projects semantically similar images together. 
}

To explore a large dataset effectively, an OoD-based sampling method was developed to build a hierarchy. 
\yafeng{The top level has the global view and each zoom-in moves to a lower level in the hierarchy for local exploration.}
The basic idea of this method is to probabilistically under-sample dense regions with fewer OoD \changjian{samples} and over-sample sparse regions with more OoD \changjian{samples}. 
It thus preserves both the overall data distribution and \changjian{OoD samples}. 
Accordingly, the grid-based visualization\shixia{, together with} this sampling technique, facilitates the examination of OoD \changjian{samples} among the normal samples at different levels of detail (\textbf{R1}).
In addition, a rich set of interactions is provided to help explore the OoD context. 
For example, a hybrid comparison, including juxtaposition and superposition~\cite{gleicher2011visual,gleicher2018considerations}, 
was developed to enable the side-by-side comparison between the training and test datasets and  the comparison of the \changjian{OoD sample(s)} of interest with
relevant training samples (\textbf{R2}). \looseness=-1

\changjian{With the help of visualization, the user can identify the reason why OoD samples appear.
Accordingly, training data will be expanded to start a new round of analysis (Fig.~\ref{fig:analysis_pipeline}(d)).}

%% file: data-modeling.tex
\section{Ensemble OoD Detection Method}\label{sec:detection}

Here, we introduce the method for detecting OoD samples. 

\myparagraph{Basic idea.}
Deep ensembles has been shown to \docsecondrev{perform best at OoD detection}~\cite{lakshminarayanan2017simple}. 
Thus, our detection method is based on it.
In particular, we thoroughly study what kind of combination in the ensembles can boost performance.
We find that, theoretically, if all aspects of the data distribution have been well covered by the model (\emph{model capacity} for short), the best performance will be achieved~\cite{hoeting1999bayesian}.
Generally, the coverage of the employed features and the algorithms are the major factors that influence the model capacity~\cite{lakshminarayanan2017simple}. 
Therefore, we 
\yafeng{try}
to improve the model capacity by enlarging the feature set and the algorithm set.

\myparagraph{Enlarging the feature set.}
For the feature set, deep ensembles covers a set of high-level features (learned by neural networks). 
It builds all the possible matches between a single classification algorithm~\cite{goodfellow2016deep} and these features. 
In contrast, we enlarge the feature set by including both low-level features, such as SIFT~\cite{lowe2004distinctive}, superpixels~\cite{lakkaraju2017identifying}, ORB~\cite{rublee2011orb}, and BRIEF~\cite{calonder2010brief}, 
and high-level features.
With a more comprehensive feature set, the developed ensemble model covers more aspects of the sample distribution.

\myparagraph{Enlarging the algorithm set.}
For the algorithm set, deep ensembles uses a softmax classifier with learned parameters\shixia{, which may fail to cover the whole model space and thus reach a local optimum}.
\shixia{To compensate for this,}
we use the same classification algorithm as \doc{the} base and build an algorithm family by varying the hyper-parameters in the feasible space, 
with the goal of better approximating the model space, and \doc{thereby improving} the model capacity.\looseness= -1 

In our implementation, the algorithm family is derived from 
logistic regression by varying its hyper-parameter (regularization coefficient).
\changjian{
A series of classifiers in our ensemble method \gramsecondrev{is} thus created by combining each feature with each logistic regression model, each of which gives a sample a probability distribution over classes. 
The OoD score of each sample is the entropy of the averaged probability distribution of all classifiers~\cite{lakshminarayanan2017simple}.}
\doc{Samples} with higher \changjian{OoD} 
scores are OoD \changjian{samples}. \looseness=-1
\shixia{The experiment in Sec.~\ref{subsec:ooddetect} shows that our method performs very well even with a simple algorithm such as logistic regression}.

%% file: visualization.tex
\section{Grid-Based OoD Visualization}\label{sec:vis}   

\begin{figure*}[t]
  \centering
  \includegraphics[width =\linewidth]{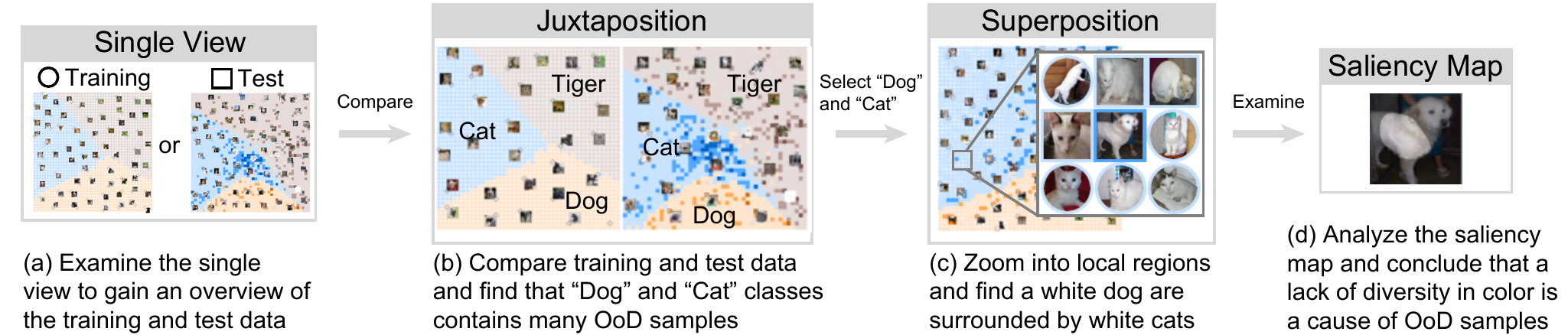}
    \caption{A typical \changjian{analysis} workflow of OoDAnalyzer.
    }
    \label{fig:workflow} 
    \vspace{-5mm}
\end{figure*}

\subsection{Overview}
Since the OoD detection method does not explain the detection decision, 
we \docsecondrev{have developed} a visualization for producing visual explanations that help users understand why.
\shixia{It consists of two major components: 1) a sample view to illustrate the distributions of the training/test data and compare them side-by-side; 2) an explanation view to disclose the reason why OoD samples appear.
}

\myparagraph{Sample distribution as grid.}
To facilitate the analysis of OoD samples, \docsecondrev{experts need to analyze} them in the context of other samples. 
We initially project data on a 2D plane as scattered points using dimension reduction techniques, such as t-SNE~\cite{tsne}. 
Despite the advantages of cluster separation and data distribution preservation, the irregular distribution of scattered points does not adequately support sample-based exploration due to the issues of overlap and space-wasting, especially when 2D points are rendered as the corresponding images.
Recent studies have shown that grid layouts are effective \docsecondrev{at} 
providing \docsecondrev{a} content overview of data, such as images~\cite{schoeffmann2011similarity, fried2015isomatch}, grouped networks~\cite{yoghourdjian2015high, marriott2012memorability}, and keywords~\cite{li2005grid, kojima2008fast}.
As a result, we propose \doc{using} 2D grid layouts to visualize the images and their similarity relationships (Fig.~\ref{fig:teaser}(c)).
Each grid cell corresponds to a data sample\gramsecondrev{,} and rectangular images can be displayed with minimal \gramsecondrev{space-wasting} and zero-overlap. \looseness=-1


\myparagraph{Prediction explanation as saliency map.}
To help users understand the underlying model prediction for a specific image,
we employ a saliency map developed by Selvaraju \etal~\cite{selvaraju2017grad} to highlight the important regions in the image for predicting the class concept.
\changjian{The visual explanation conveyed in this map is helpful for identifying the root cause of the appearance of OoD samples.}
For example, in Fig.~\ref{fig:analysis_pipeline}C, 
the white dog is predicted \docsecondrev{to be} a cat because the body (\doc{white in color}) is an important region \shixia{for} prediction. 
In normal cases, the dog head is usually the key \doc{to} making the corresponding decision. 
\changjian{After identifying many similar cases like this, we believe that a lack of diversity in color is a major cause 
\yafeng{of}
these OoD samples.}
\looseness=-1

\myparagraph{Analysis workflow}.
The two \docsecondrev{major} components \docsecondrev{of the visualization}, along with the developed controls and visualization aids (Fig.~\ref{fig:teaser}(c)), create an interactive OoD analysis environment.   
A typical analysis workflow is illustrated in Fig.~\ref{fig:workflow}. 
After loading training and test data, a user can first examine the single dataset to \doc{gain} an overview of the data distribution (Fig.~\ref{fig:workflow}(a)).
S/he can switch between training \yafeng{and} test data.
To compare the distribution of training and test data \changjian{side-by-side}, 
s/he can then use the juxtaposition comparison mode to identify which categories contain OoD \changjian{samples}
and decide if s/he wants to select a few categories to explore (Fig.~\ref{fig:workflow}(b)).
After selecting categories of interest, 
s/he can switch to the superposition mode to explore the OoD \changjian{samples} in the context of relevant training samples (Fig.~\ref{fig:workflow}(c)). 
Instead of using the juxtaposition mode first, if the user finds interesting OoD regions in the single view, s/he can directly use the superposition mode to explore the OoD \changjian{samples}.
In any mode,
the user can zoom into a region with OoD \changjian{samples} or inspect the original images of selected samples as well as their saliency maps (Fig.~\ref{fig:workflow}(d)).
With the help of saliency maps, 
\doc{the user can better understand why these samples are OoD samples.} \looseness=-1

\subsection{\texorpdfstring{$k$}{k}NN-based Grid Layout}
\label{subsec:knn-appro}

\myparagraph{Grid Layout.}
A practical \docsecondrev{method for generating a} grid layout mapping is \docsecondrev{the} two-step approach described in~\cite{fried2015isomatch,Markovtsev2017}: 
the data is first projected as a set of 2D scattered points \doc{using} dimension reduction techniques; then these 2D points are assigned to grid points by solving a linear assignment problem.  
\vspace{0mm}
\begin{itemize}[leftmargin=*]\setlength\itemsep{0mm}

 \item[--] \textbf{Step 1: Dimension reduction.} Assume that each instance in the dataset is associated with a feature vector. 
 This method projects all the feature vectors onto a 2D plane via Isomap~\cite{isomap} or t-SNE~\cite{tsne}.  
In our implementation, we used t-SNE.
For convenience, we denote the projected points by $\mx_i \in \mathbb{R}^2$. 
  \item[--] \textbf{Step 2: Linear assignment.}  A 2D $m \times n$ grid  is created over the bounding box of projected points. \changjian{Each grid point is represented by its center $\my_j \in \mathbb{R}^2$.} Without loss of generality, assume that the size ($N$) of instances in the dataset is equal to $m \times n$. This assumption can be easily enforced by adding ``dummy instances.''  An assignment from $X = \{\mx_i\}_{i=1}^N$ to $Y = \{\my_i\}_{i=1}^N$ can be denoted by $\delta_{ij}$ where $\delta_{ij}=1$ means that the $i$-th instance $\mx_i$ is assigned to the $j$-th grid point $\my_j$; otherwise $\delta_{ij}=0$. Let the assignment cost from $\mx_i$ to $\my_j$ be $w_{ij} = \|\mx_i - \my_j\|_2$, so that the optimal assignment can be achieved by minimizing the total assignment cost:
\begin{equation} \label{eq:lap}
\vspace{-2mm}
\begin{aligned}
& \underset{\delta_{ij}}{\textbf{minimize}}
 &                                                    & {\textstyle\sum_{i=1}^N \sum_{j=1}^N} w_{ij}\delta_{ij}                                                                    \\
& \text{subject to}
&                                                    &  {\textstyle\sum_{i=1}^N}{\delta_{ij}} = 1,          \forall  j \in \{1, \ldots, N\};                     \\
&                          &                           &   {\textstyle\sum_{j=1}^N}{\delta_{ij}} = 1,           \forall i \in \{1, \ldots, N\};                             \\
&  & & \delta_{ij} \in \{0,1\}, \forall i, j.
\end{aligned}
\end{equation}
The constraints enforce that \docsecondrev{the} assignment is bijective, \ie  each instance has a unique assignment in the grid, and vice versa. 
\end{itemize}

The above linear assignment problem is equivalent to a weighted bipartite graph matching problem on a weighted complete graph $G=(X, Y, E)$ by regarding $X$ and $Y$ as two independent sets, $E$ as the set of edges between $X$ and $Y$, and $w_{ij}$ as the weight defined on edge $\overline{\mx_i\my_j}$. 
The matching problem (Eq.~\ref{eq:lap}) has been extensively studied \doc{previously}~\cite{jonker1987shortest, munkres1957algorithms}, and the Jonker-Volgenant (JV) algorithm~\cite{jonker1987shortest} has an overall satisfactory and stable time performance~\cite{dell2000algorithms}.
The JV algorithm consists of $N$ path augmenting processes, 
in each of which an $O(N^2)$ Dijkstra algorithm is executed to find the augmenting path with the \docsecondrev{shortest} distance. 
Accordingly, the time complexity of the JV algorithm is $O(N^3)$. 
For our interactive visualization,  as instances are resampled during \doc{user} exploration, the grid layout must generate at interactive rates (less than 1 second for a thousand instances), but the JV algorithm cannot meet this requirement.

\myparagraph{$k$NN-based bipartite graph matching}.
To generate the grid layout at interactive rates, we \docsecondrev{have developed} a $k$NN-based bipartite graph matching method to speed up the layout process while
achieving a satisfactory approximation \docsecondrev{of} the optimal result.
\docsecondrev{Since} most instances will be assigned to grid points near their projection location, 
a simple idea \docsecondrev{for acceleration}
is \docsecondrev{pruning} the edges
\docsecondrev{that connect} instances with distant candidate grid points.
Specifically, \changjian{we reduce} the number of candidate grid points of every $\mx_i$ from $N$ to a smaller number $k$ ($k < N$),
\ie for each $\mx_i$, \docsecondrev{we enforce} its $N-k$ related $\delta_{ij}$s to be $0$.
\shixia{As the Dijkstra algorithm checks all vertices once and traverses their neighbors during the checks~\cite{bondy1976graph, leiserson2001introduction}, 
\yafeng{its }time complexity 
is $O(kN)$ if each vertex only has k neighbors.}
\changjian{Thus the time complexity} of the JV algorithm is reduced to $O(kN^2)$. If $k$ is much smaller than $N$, the time complexity can be
regarded as $O(N^2)$. However, a na\"{i}ve reduction cannot guarantee that 
there \docsecondrev{will be} a solution that fulfills all the constraints of Eq.~(\ref{eq:lap}) because it may violate the sufficient and necessary condition of graph matching, which is known as \emph{Hall's theorem}~\cite{bondy1976graph}. \looseness=-1

\begin{theorem}[Hall's theorem]
A bipartite graph $G=(X,Y,E)$ has a matching if and only if the number of elements of any subset $S$ of $X$ is less than or equal to the number of elements of its corresponding adjacent set $\Gamma(S)$, namely, 
\begin{equation*}
\begin{array}{ccc}
|S| \leq |\Gamma(S)|, & & \forall S \subseteq X.  
\end{array}
\end{equation*}
\end{theorem}
\changjian{$\Gamma(S)$ is the set of all neighbors of the vertices in $S$.}
\begin{equation*}
    \Gamma(S) = \{ \my \in Y: \forall \mx \in S, \ \overline{\mx \my} \in E \} \subset Y
\end{equation*}

However, it is difficult to use Hall's theorem to check whether \docsecondrev{a} matching exists because it is an NP-hard problem due to the combinatorics of subsets.
Instead, we use a sufficient condition derived from Hall's theorem, also known as \cjsecondrev{the} \emph{marriage theorem}~\cite{bondy1976graph}.
\newtheorem{corollary}{Corollary}
\begin{corollary}[Marriage theorem]
A bipartite graph $G=(X,Y,E)$ has a matching if the degree of each vertex $\mx \in X$ equals $k$ and the degree of each vertex $\my \in Y$ also equals $k$. $k$ is a natural number.
\end{corollary}

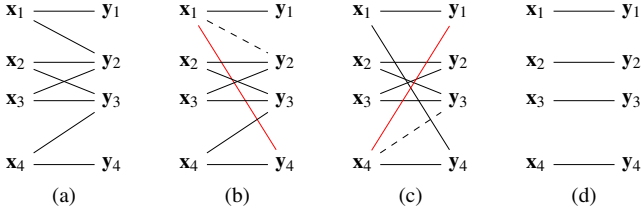
\begin{figure}[t]
\vspace{-3mm}
\centerline{
\scalebox{0.85}{
\begin{tikzpicture}
  \node at (0,2.4)  (x1) {$\mx_1$};
  \node at (1.5,2.4) (y1) {$\my_1$};
  \node at (0,1.6) (x2) {$\mx_2$};
  \node at (1.5,1.6) (y2){$\my_2$};
  \node at (0,1) (x3) {$\mx_3$};
  \node at (1.5,1) (y3) {$\my_3$};
  \node at (0,0) (x4) {$\mx_4$};
  \node at (1.5,0) (y4) {$\my_4$};
  \draw (x1) -- (y1);   \draw (x1) -- (y2);  
   \draw (x2) -- (y2);   \draw (x2) -- (y3);   
  \draw (x3) -- (y2);  \draw (x3) -- (y3); 
  \draw (x4) -- (y3); \draw (x4) -- (y4);
   \node at (0.75,-0.5) (a) {(a)};
\end{tikzpicture}}
 \hfill
 \scalebox{0.85}{
\begin{tikzpicture}
  \node at (0,2.4)  (x1) {$\mx_1$};
  \node at (1.5,2.4) (y1) {$\my_1$};
  \node at (0,1.6) (x2) {$\mx_2$};
  \node at (1.5,1.6) (y2){$\my_2$};
  \node at (0,1) (x3) {$\mx_3$};
  \node at (1.5,1) (y3) {$\my_3$};
  \node at (0,0) (x4) {$\mx_4$};
  \node at (1.5,0) (y4) {$\my_4$};
  \draw (x1) -- (y1);   \draw[dashed] (x1) -- (y2);  \draw[red] (x1) -- (y4);
   \draw (x2) -- (y2);   \draw (x2) -- (y3);
  \draw (x3) -- (y2);  \draw (x3) -- (y3);
  \draw (x4) -- (y3); \draw (x4) -- (y4);
   \node at (0.75,-0.5) (b) {(b)};
\end{tikzpicture}}
 \hfill
 \scalebox{0.85}{
\begin{tikzpicture}
  \node at (0,2.4)  (x1) {$\mx_1$};
  \node at (1.5,2.4) (y1) {$\my_1$};
  \node at (0,1.6) (x2) {$\mx_2$};
  \node at (1.5,1.6) (y2){$\my_2$};
  \node at (0,1) (x3) {$\mx_3$};
  \node at (1.5,1) (y3) {$\my_3$};
  \node at (0,0) (x4) {$\mx_4$};
  \node at (1.5,0) (y4) {$\my_4$};
 \draw (x1) -- (y1);     \draw (x1) -- (y4);
   \draw (x2) -- (y2);   \draw (x2) -- (y3);
  \draw (x3) -- (y2);  \draw (x3) -- (y3);
 \draw[red] (x4) -- (y1);  \draw [dashed]  (x4) -- (y3); \draw (x4) -- (y4);
  \node at (0.75,-0.5) (c) {(c)};
\end{tikzpicture}}
\hfill
\scalebox{0.85}{
\begin{tikzpicture}
  \node at (0,2.4)  (x1) {$\mx_1$};
  \node at (1.5,2.4) (y1) {$\my_1$};
  \node at (0,1.6) (x2) {$\mx_2$};
  \node at (1.5,1.6) (y2){$\my_2$};
  \node at (0,1) (x3) {$\mx_3$};
  \node at (1.5,1) (y3) {$\my_3$};
  \node at (0,0) (x4) {$\mx_4$};
  \node at (1.5,0) (y4) {$\my_4$};
 \draw (x1) -- (y1);   
   \draw (x2) -- (y2);   
\draw (x3) -- (y3);
 \draw (x4) -- (y4);
 \node at (0.75,-0.5) (d) {(d)};
\end{tikzpicture}
}}
    \caption{$k$NN-based bipartite graph matching on a simple example.  (a) The $k$NN-based subgraph initialization with $k=2$; (b,c) intermediate and final results of greedy modification. The dashed edges are removed and the red edges are added; (d) the bipartite graph matching obtained by applying the JV algorithm to (c).  }\label{fig:greedy-method} 
\end{figure}

Based on this corollary, we propose a two-step method to construct a matching-existed bipartite graph.
\begin{itemize}[leftmargin=*]\setlength\itemsep{1mm}
\item[--] \textbf{Step 1: $k$NN-based subgraph initialization}. 
\docsecondrev{First,} an empty edge set $E_r$ is created. 
Then for each $\mx \in X$, we select its $k$-nearest neighbors in $Y$ and add edges between $\mx$ and each of its neighbors into $E_r$.  In this way\gramsecondrev{,} an initial bipartite graph $G_r = (X, Y, E_r)$ is created. We use $\md(\cdot)$ to denote the degree of a vertex in the graph. 
The subscripts $`>`, `=`, `<`$ denote the vertex set whose degrees are larger than $k$, equal to $k$, or less than $k$.
For instance,
$X$ is divided into three subgroups $X_> := \{\mx \in X | \deg(\mx)>k\}$, $X_= := \{\mx \in X | \deg(\mx)=k\}$, $X_< := \{\mx \in X | \deg(\mx) < k\}$. 
After the initialization, $X_< = \varnothing, X_> = \varnothing$. \looseness=-1
\item[--] \textbf{Step 2: Greedy subgraph modification.}  The initial graph $G_r$ may not satisfy the marriage theorem because $Y_>$ and $Y_<$ can be nonempty. 
We apply a greedy graph modification to make them empty, while keeping $X_>$ and $X_<$ empty, too.
We examine all the vertices in $Y_>$ in descending order prioritized by the vertex degree and the sum of their edge weights in $E_r$. For each examined vertex $\my \in Y_>$, we scan all of its graph edges in descending order prioritized by edge weights: for the scanned edge $\overline{\mx\my}$, we propose
to delete it, however, since $\md(\mx) = k$, the edge deletion will move $\mx$ into $X_<$ so that $X_<$ is not empty. To guarantee that $X_<$ is always empty, 
\docsecondrev{we pick the edge $\overline{\mx\my^\prime}$ with the smallest edge weight from the edge set $\{\overline{\mx\my^\prime}: \overline{\mx\my^\prime} \notin E_r, \my^\prime \in Y_< \}$, 
and add it to} $E_r$ while deleting $\overline{\mx\my}$. 
In case $U_\mx$ is empty and $\my^\prime$ does not exist, we scan the next edge of $\my$; otherwise, we stop the edge scan for $\my$. 
The greedy modification is executed until $Y_> = \varnothing, Y_< = \varnothing$, and the final graph $G_r$ satisfies the marriage theorem, \ie $X = X_=, Y = Y_=$. 
\end{itemize}

In Fig.~\ref{fig:greedy-method}, we use a simple example to illustrate our $k$NN-based algorithm.
Fig.~\ref{fig:greedy-method}(a) shows an initial graph for the set of $X=\{\mx_1, \mx_2, \mx_3, \mx_4\}$ and $Y=\{\my_1, \my_2, \my_3, \my_4\}$, with $k=2$, in which the edge length represents its weight. 
Among the vertices in $Y_<$, $\my_2$ has the largest degree and the largest sum of the edge weights, 
so we pick it first and remove \docsecondrev{the} edge $\overline{\mx_1\my_2}$\docsecondrev{, which} has the largest weight. 
To ensure $\mx_1$ is still inside $X_=$, we add a new edge $\overline{\mx_1\my_4}$ ( Fig.~\ref{fig:greedy-method}(b)),
since $\my_4$ is the only vertex in $Y_<$ not \doc{connected to} $\mx_1$.
The greedy modification further removes $\overline{\mx_4\my_3}$ and adds $\overline{\mx_4\my_1}$.  The resulting graph Fig.~\ref{fig:greedy-method}(c) satisfies the marriage theorem since the degree of any vertex is $k=2$. After applying the JV algorithm on the modified graph (c), a bipartite graph matching is obtained (Fig.~\ref{fig:greedy-method}(d)). \looseness=-1

A formal proof of the correctness of the greedy modification is given in \docsecondrev{the} supplemental material. 
As each edge in $E_r$ is checked at most once,
the worst time complexity of the greedy modification is $O(kN)$, 
where $kN$ is the size of $E_r$. 
Once \docsecondrev{a} matching-existed bipartite graph $G_r=(X,Y,E_r)$ is created, we apply the JV algorithm to compute the grid layout.  
Our method provides an approximate solution to Eq.~\ref{eq:lap} with high efficiency. In Sec.~\ref{subsec:knneval}, we discuss the practical choice of $k$.

\subsection{Interactive Exploration}\label{subsec:interact}
To facilitate the exploratory analysis of OoD \changjian{samples}, three interactions, comparison, OoD recommendation, and OoD-based zooming, are provided.

\myparagraph{Comparison}. 
OoDAnalyzer provides two 
comparison strategies, juxtaposition and superposition~\cite{gleicher2011visual,gleicher2018considerations}, 
to support the detailed examination of \changjian{OoD samples} in different contexts.
We use circles and squares to represent training and test samples, respectively (Fig.~\ref{fig:teaser}E).
The juxtaposition design places the grid-based visualizations of the training and test datasets side-by-side (Fig.~\ref{fig:analysis_pipeline}A), \yafeng{and} 
the user can 
identify which categories contain OoD \changjian{samples}.
After selecting categories of interest, 
the superposition design, which presents the training and test datasets in the same grid-based visualization (Fig.~\ref{fig:analysis_pipeline}B), is utilized to visualize OoD \changjian{samples} in the context of relevant training samples. 
\doc{After} a detailed comparison of the \changjian{OoD samples with their} relevant training samples, 
as well as \doc{a} careful examination of the saliency maps, 
\doc{users can} gradually dig out the underlying reasons 
for \changjian{their appearance}.\looseness=-1

\myparagraph{OoD recommendation}. 
To provide an informative start for \docsecondrev{OoD} analysis, the ensemble algorithm recommends a set of OoD \changjian{samples} for further analysis in the grid-based visualization.
To help better disclose the \changjian{OoD samples detected by the ensemble model and their categories estimated by the prediction model}, 
a combination of qualitative and sequential coloring is used to encode OoD \changjian{samples} in different categories (Fig.~\ref{fig:teaser}(b)). 
In particular, the color of the grid \changjian{cell and border} encodes the category, and a sequential color scheme with 3 colors is used to represent the OoD scores of the samples.
The darker the \docsecondrev{color}, the higher the OoD \docsecondrev{score}. 
The lightest color represents the normal samples in each category.
The user can select the OoD \changjian{samples} with darker borders for further analysis.  \looseness=-1

\begin{figure}[t]
  \centering
  \includegraphics[width =\linewidth]{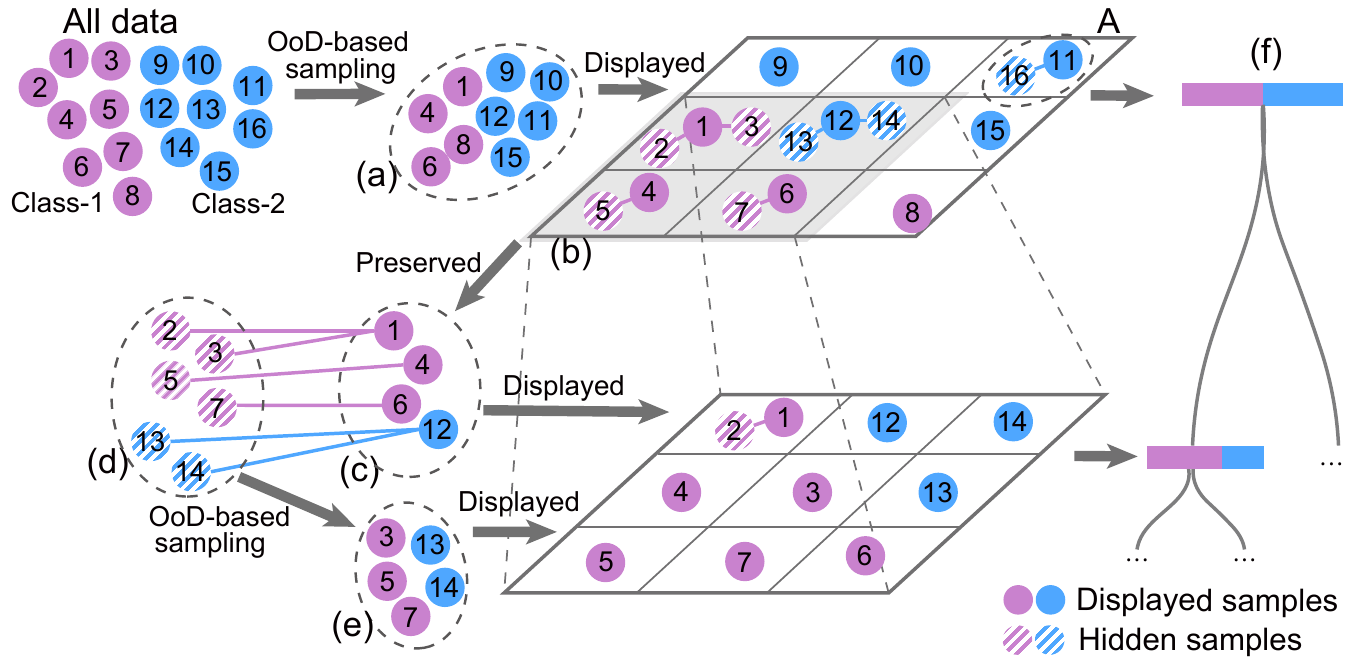}
    \caption{\changjian{Illustration of OoD-based zooming. (a) Initial sampling result; (b) A sub-area (gray) selected by the user; (c) 4 displayed samples preserved to the next level; (d) 6 hidden samples assigned to these 4 displayed samples; (e) 5 samples \docsecondrev{selected} from the 6 hidden samples; (f) Hierarchy. }
    }
    \label{fig:sampling} 
    \vspace{-3mm}
\end{figure}

\myparagraph{OoD-based zooming}.
To reduce visual clutter caused by \gramsecondrev{a} large number of samples, 
we \docsecondrev{employ} an OoD-based sampling technique 
inspired by \docsecondrev{outlier} biased sampling~\cite{xiang2019interactive}. 
It probabilistically under-samples dense regions with fewer OoD samples and over-samples sparse regions with more OoD samples. 
This strategy 
\yafeng{balances}
providing more OoD samples and maintaining \cjsecondrev{the} overall data distribution.
\yafeng{To navigate between \doc{a} global overview and local details, }
we use the OoD-based sampling technique to dynamically build a hierarchy structure (Fig.~\ref{fig:teaser}I and Fig.~\ref{fig:sampling}(f)). 

\shixia{Fig.~\ref{fig:sampling} uses a simple example to illustrate the basic idea of the sampling-based hierarchy building method. 
Initially, $3 \times 3$ (the max number of grids) samples are sampled 
and displayed in the grids (Fig.~\ref{fig:sampling}(a)). 
Other samples are 
\yafeng{hidden}
and virtually assigned to their nearest displayed samples (Fig.~\ref{fig:sampling}A).}
The user can select a sub-area (Fig.~\ref{fig:sampling}(b)) to zoom into a subset of the data, which contains both 
the displayed samples (Fig.~\ref{fig:sampling}(c)) at the current level and the hidden samples (Fig.~\ref{fig:sampling}(d)) assigned to them in this selected area.
\changjian{If the size of the subset (denoted as $V$) is larger than $3 \times 3$, }
$3 \times 3$ samples are \changjian{obtained} from this subset, which contains all the displayed samples to maintain the mental map 
\changjian{and the samples \docsecondrev{selected} from the hidden samples (Fig.~\ref{fig:sampling}(e))}.
\changjian{Otherwise,}
a $R \times R$ ($R \leq 3$) grid layout is created and all samples of this subset are displayed in the grids.
Here $R$ is the smallest positive integer satisfying $R \times R \geq V$. 
To track user actions, a hierarchy (Fig.~\ref{fig:sampling}(f)) is constructed.
The root node represents all samples.
Each time a subset is selected from a node, 
\yafeng{a child node is created to represent this selected subset.}
The node consists of several color bars where the color indicates the data category and the bar length encodes the number of samples in the category. 
\changjian{The user can} right click on a particular node to load the displayed samples under that node and then continue to select another sub-area for detailed examination.
With such interactions, the user can examine the samples at different levels of \docsecondrev{detail}. 

\changjian{If the grid size is larger than a specified threshold, all images are displayed in grids.
Otherwise, we only show dozens of representative images with higher OoD scores. 
Each image is placed near the center of the grid it belongs to (Fig.~\ref{fig:teaser}D). 
The images are placed one by one in descending order of their OoD scores.
\doc{An} image is not placed if it is too close to previously placed images.} \looseness=-1



%% file: application.tex
\section{Evaluation}\label{sec:eval}
\label{sec:evaluation}
To validate our approach, we performed a quantitative evaluation of the ensemble OoD detection algorithm and \cjsecondrev{the} $k$NN-based grid layout algorithm (Sec.~\ref{subsec:eval}).
Two case studies were conducted to demonstrate how OoDAnalyzer can be used to effectively identify OoD samples in test data (Sec.~\ref{subsec:case}).  All 
experiments were run on a desktop computer with an Intel Xeon E5-2630 CPU (2.2GHz) and 128 GB of memory. \looseness=-1

\subsection{Quantitative Evaluation}\label{subsec:eval}
\subsubsection{OoD Detection}\label{subsec:ooddetect}
We evaluated the ensemble OoD detection algorithm on classification tasks. 
Three datasets were tested, 
\yafeng{and}
the ground-truth OoD samples were given for evaluating the detection accuracy. 
(1) \textbf{Dog\&Cat}: \docsecondrev{this} training set has 2,603 images, including various dark-colored dogs and light-colored cats.  The test dataset has 637 normal images and 640 OoD images (light-colored dogs and dark-colored cats).
All the data \doc{come} from a Kaggle competition~\cite{kaggle}. 
The classification task is to distinguish between dogs and cats.
(2) \textbf{SVHN-3-5}: 5,768 images containing dark digits 3 on light backgrounds 
or light digits 5 on 
dark backgrounds \docsecondrev{were} selected from the Street View House Numbers (SVHN) dataset~\cite{netzer2011reading} as the training data. 
The test dataset includes 2,668 normal images and 2,578 OoD images
where the color of \docsecondrev{the} digits and their background \docsecondrev{have been} reversed to that of training data. 
The classification task is to distinguish between \docsecondrev{the digits} 3 and 5.
(3) \textbf{MNIST}: we use the original training dataset (60,000 images) and the test dataset (10,000 images) of MNIST~\cite{lecun1998gradient} 
as our training set and normal test \docsecondrev{set}, respectively. 
18,726 images from the NotMNIST dataset~\cite{notmnist} \docsecondrev{were} treated as OoD samples. 
The classification task is the standard 10-class handwritten digit classification. \looseness=-1

We measured the quality of OoD detection by:

\begin{itemize}[leftmargin=*]\setlength\itemsep{0mm}
\item[-] \textbf{AUROC} \shixia{is the area under the receiver operating characteristic curve, which measures how well a classifier discriminates between two classes of \gramsecondrev{the} outcome}.
\item[-] \textbf{AUPR} \shixia{is the area under the precision-recall curve, which is used to evaluate the global accuracy of a classifier}. 
\item[-] \shixia{\textbf{Top-$K$ precision ($\mathrm{Prec}_K$)} is the fraction of OoD samples among the first $K$ samples with the highest OoD scores. 
A higher Top-$K$ precision \doc{means} a better OoD detection performance.}

\end{itemize}

The feature set and the algorithm set \docsecondrev{can both} impact the OoD detection results. 
For the feature set, the numbers of the high-level and low-level features are determined based on \doc{a review of the literature} and our experiments. 
Deep ensembles~\cite{lakshminarayanan2017simple} suggests that a set \doc{with 5 or more} high-level features performs well. 
Adding more high-level features would \doc{lead to better performance} but also increase the training time.
Therefore, we used 6 high-level features extracted by 3 VGG nets with random initial parameters, Inception-v3 net, MobileNet, and ResNet.
A recent study has indicated that SIFT, BRIEF, and ORB are widely-used local image feature descriptors~\cite{zhang2019learning}. 
In addition, superpixel was used to detect OoD samples in a recent work~\cite{lakkaraju2017identifying}.  
As a result, we employ\changjian{ed} these \changjian{4} low-level features. 

\setlength{\tabcolsep}{5.5pt}
\begin{table}[t]
\centering
\caption{Comparison of OoD detection algorithms (\small{larger values are better}). }
 \scalebox{0.93}{
\begin{tabular}{c|ccccc}
 \toprule
   & AUROC $\uparrow$ & AUPR $\uparrow$ & $\mathrm{Prec}_{50}$ $\uparrow$ & $\mathrm{Prec}_{100}$ $\uparrow$ & $\mathrm{Prec}_{200}$ $\uparrow$ \\
   \midrule \textbf{Dog\&Cat} & \multicolumn{4}{c}{}   \\
 deep ensembles
&0.6518&0.6065&0.680&0.680& 0.685\\
 S-OoD & 0.8076 & 0.8170 & 0.900 & 0.920 & \textbf{0.935}\\
 M-OoD & \textbf{0.8107} & \textbf{0.8172} & \textbf{0.920} & \textbf{0.930} & 0.930 \\
 \midrule \textbf{SVHN-3-5}  & \multicolumn{4}{c}{} \\
 deep ensembles&0.4976&0.5114&0.600&0.580&0.545 \\
 S-OoD&0.8770&0.8404&0.800&0.830&0.805 \\
 M-OoD&\textbf{0.9060}&\textbf{0.9062}&\textbf{0.980}&\textbf{0.950}&\textbf{0.945} \\
  \midrule \textbf{MNIST}  & \multicolumn{4}{c}{} \\
 deep ensembles&0.9908&0.9883&\textbf{1.000}&0.990&0.995 \\
 S-OoD & 0.9954 & 0.9946 & \textbf{1.000} & \textbf{1.000} & \textbf{1.000} \\
 M-OoD & \textbf{0.9969} & \textbf{0.9966} & \textbf{1.000} & \textbf{1.000} & \textbf{1.000} \\
 \bottomrule
\end{tabular}
 } 
\label{tab:ood-benchmark} \vspace{-2mm}
\end{table}
\setlength{\tabcolsep}{6pt}

\setlength{\tabcolsep}{5pt}
\begin{table}[b]
\centering
\caption{The computation time $\mathrm{T}$ of our $k$NN-based grid layout method on real data. ``BL'' represents the baseline method that obtains optimal results but with high computational cost, $O(N^3)$. Here time is measured in seconds.}
\begin{tabular}{c|cccccc}
 \toprule
  & BL & $k=50$ & $k=100$ & $k=200$ & $k=500$ & $k=1000$ \\
 \midrule
 REA & 4.90 & 0.52 & 0.63 & 0.78 & 1.19 & 1.77 \\
 Animals & 3.43 & 0.54 & 0.59 & 0.72 & 1.07 & 1.57 \\
 SVHN & 3.65 & 0.53 & 0.58 & 0.72 & 1.09 & 1.60 \\
 CIFAR10 & 2.90 & 0.51 & 0.59 & 0.69 & 0.98 & 1.37 \\
 MNIST & 3.51 & 0.62 & 0.62 & 1.01 & 1.44 & 1.44 \\
 \bottomrule
\end{tabular} 
\label{tab:real-timing} 
\end{table}
\setlength{\tabcolsep}{6pt}

\begin{table}[b]
\centering
\caption{The approximation quality $\mathrm{C}_r$ of our $k$NN-based grid layout algorithm on real data. Lower values are better.}
\begin{tabular}{c|cccccc}
 \toprule
  & $k=50$ & $k=100$ & $k=200$ & $k=500$ & $k=1000$ \\
 \midrule
 REA & 3.49e-03 & 6.49e-04 & 7.04e-05 & 3.17e-06 & 4.30e-07 \\
 Animals & 1.00e-03 & 2.74e-04 & 4.33e-05 & 1.28e-06 & 2.03e-07 \\
 SVHN & 1.96e-03 & 8.31e-04 & 1.14e-04 & 7.99e-08 & 2.69e-10 \\
 CIFAR10 & 2.19e-03 & 9.56e-04 & 4.34e-05 & 1.28e-07 & 5.07e-10 \\
 MNIST & 1.18e-03 & 2.48e-04 & 4.19e-05 & 5.85e-07 & 1.69e-07 \\
 \bottomrule
\end{tabular}
\label{tab:real-approx} \vspace{-3mm}
\end{table}

For the algorithm set, we employed logistic regression. 
This was inspired by the design of deep neural networks for classification~\cite{goodfellow2016deep}. 
It uses softmax regression\docsecondrev{, which} is equivalent to logistic regression for binary classification. 
\changjian{The previous} study has shown that the regularization coefficient of logistic regression is generally chosen from an exponentially growing sequence~\cite{scikit-learn}. 
As a result, the regularization coefficient \cjsecondrev{was} chosen from \{$10^k$: $k = -5,\cdots,-1,0, 1,\cdots,5$\}. 
To better cover the hyper-parameter space, we \cjsecondrev{tried} to 
select \docsecondrev{values} with the largest overall distance among the regularization coefficients of the logistic regression algorithms in each ensemble method. 
For example, when the classifier number in the ensemble method is 3, we set the regularization coefficient as \{$10^{-5}, 1, 10^5$\}.
In particular, we set the regularization coefficient as 1 
when the classifier number in the ensemble method \cjsecondrev{was} 1. 
We then evaluated the ensemble OoD detection method with the classifier number changing from 1 to 11. 
The experiments in \doc{the} supplemental material show that when the classifier number \docsecondrev{was} greater than or equal to 3, the ensemble OoD detection method \docsecondrev{achieved} an acceptable result. 
Thus, we set the classifier number as 3 and the regularization coefficient\changjian{s as \{$10^{-5}$, $1$, $10^{5}$\}.} 
Each of the 10 features was combined with each of the 3 logistic regression models, \doc{resulting} in 30 classifiers.
As proposed in deep ensembles~\cite{lakshminarayanan2017simple}, the entropy of the averaged predicted probabilities of the 30 classifiers was used to characterize the possibility of being an OoD sample. 
We also designed a variant of our method that only used a single logistic regression model whose regularization coefficient was 1.
These two versions are named M-OoD and S-OoD.

On the three datasets, we tested M-OoD and S-OoD, and compared them with a state-of-the-art ensemble-based method, deep ensembles~\cite{lakshminarayanan2017simple}. 
The $M$ classifiers used in deep ensembles \doc{were} derived by randomly initializing the parameters of a given neural network. 
As in~\cite{lakshminarayanan2017simple}, $M$ is set as 5. 
Table~\ref{tab:ood-benchmark} shows that (1) both M-OoD and S-OoD outperformed deep ensembles significantly on Dog\&Cat and SVHN-3-5, and slightly better on MNIST since there was not much room to improve on this dataset; (2) M-OoD predicted OoD samples more accurately than S-OoD because using a family of classifiers provides a better model capability as discussed in Sec.~\ref{sec:detection}. \looseness=-1

\subsubsection{\texorpdfstring{$k\rm {NN}$}{kNN}-based grid layout}\label{subsec:knneval}
We conducted a series of experiments to evaluate the efficiency and approximation quality of our $k$NN-based grid layout method.
In particular, we compared it with the standard two-step method, \ie the baseline method, which also uses t-SNE for dimension reduction, but applies the JV algorithm on the fully-connected bipartite graph. \looseness=-1

\myparagraph{Measures.} Two quality measures are defined for this evaluation.
\begin{itemize}[leftmargin=*]\setlength\itemsep{0mm}
\item[-] \textbf{$\mathrm{T}$}: the computational time of the bipartite matching excluding the dimension reduction.
\item[-] \textbf{$\mathrm{C}_r$}: \gramsecondrev{approximation} quality. 
\docsecondrev{We denote} $\mathrm{C}_k$ as the cost of Eq.~(\ref{eq:lap}) obtained by our $k$NN-based algorithm, 
\docsecondrev{and} $\mathrm{C}_{opt}$ as the optimal cost obtained by the baseline method. The ratio $\mathrm{C}_r := \frac{\mathrm{C}_k-\mathrm{C}_{opt}}{\mathrm{C}_{opt}}$ measures how well our method approximates the optimal solution. $\mathrm{C}_r = 0$ is optimal.
\end{itemize}

\begin{figure}[t]
  \centering
  \begin{overpic}[width=1\linewidth]{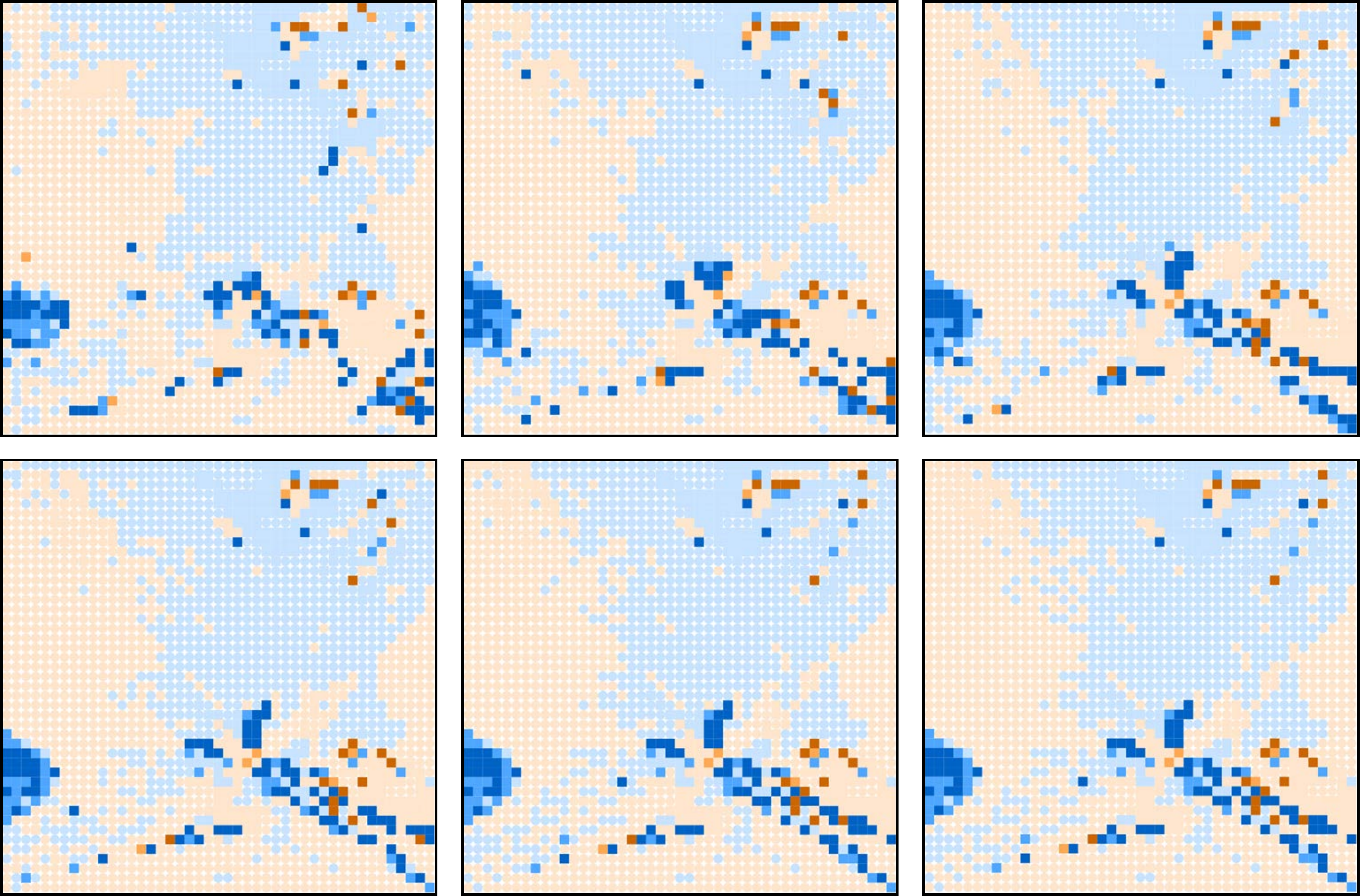}
  \put(1,62.5){\contour{white}{ $k=20$}}
  \put(35,62.5){\contour{white}{ $k=50$}}
  \put(69,62.5){\contour{white}{ $k=100$}}
  \put(1,29){\contour{white}{ $k=200$}}
  \put(35,29){\contour{white}{ $k=500$}}
  \put(69,29){\contour{white}{ Exact solution}}
  \end{overpic}
  \caption{Grid visualization of the REA dataset by \grammarly{the} $k$NN-based method with different $k$s (REA dataset). }\label{fig:varingrea}
\end{figure}

\begin{figure}[t]
  \centering
  \begin{overpic}[width=1\linewidth]{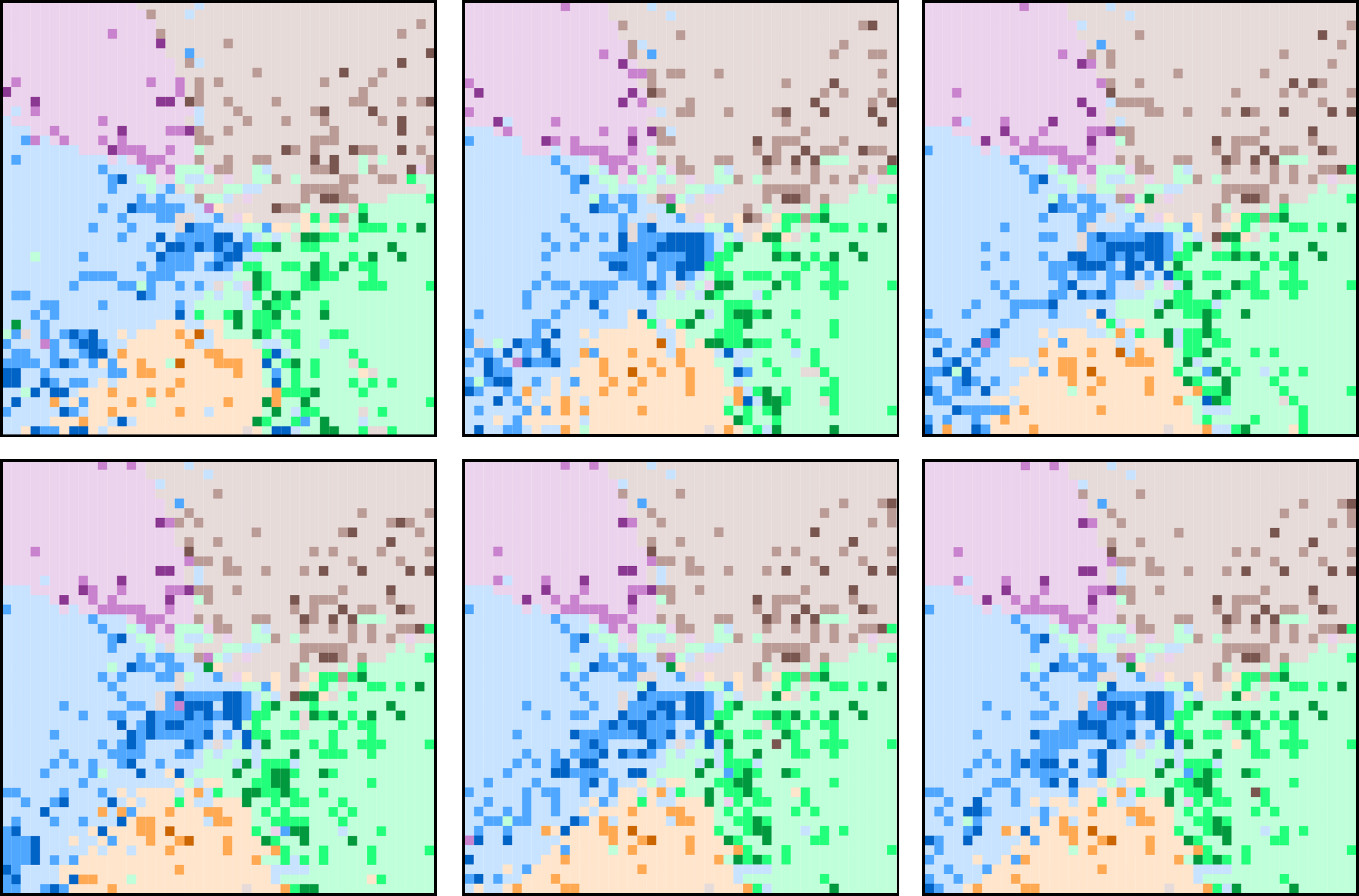}
   \put(1,62.5){\contour{white}{ $k=20$}}
  \put(35,62.5){\contour{white}{ $k=50$}}
  \put(69,62.5){\contour{white}{ $k=100$}}
  \put(1,29){\contour{white}{ $k=200$}}
  \put(35,29){\contour{white}{ $k=500$}}
  \put(69,29){\contour{white}{ Exact solution}}
   \end{overpic}
  \caption{Grid visualization of the Animal dataset by \grammarly{the} $k$NN-based method with different $k$s (animal dataset). }\label{fig:varinganimal}
  \vspace{-3mm}
\end{figure}

\myparagraph{Experiments.}  We tested the two methods on five real datasets: REA~\cite{retinal}, five animal categories (Cat, Dog, Tiger, Wolf, Rabbit) from Kaggle~\cite{kaggle} and ImageNet~\cite{russakovsky2015imagenet}, SVHN~\cite{netzer2011reading}, MNIST~\cite{lecun1998gradient}, and CIFAR10~\cite{krizhevsky2009learning}. From each dataset, we sampled 2025 ($45\times45$) images.
We utilized an open-source implementation of the JV algorithm~\cite{lapjv} in the experiments.
The results (the average values of ten trials) in terms of the quality measures are shown in Tables~\ref{tab:real-timing} and~\ref{tab:real-approx}.
The statistics of these datasets show that the cost difference ratio $\mathrm{C}_r$ is less than $1\%$ when $k\geq 50$ and 
the computational time is reduced by more than $79\%$, compared to the baseline method when $k \leq 100$.
Next, we illustrate the visual difference of the grid layouts \shixia{with different $k$s, on the REA dataset and Animal dataset}.  
Figs.~\ref{fig:varingrea} and ~\ref{fig:varinganimal} show the grid layout generated by $k$NN methods with \shixia{different values of $k$ ($k = 20, 50, 100, 200, 500$)}. 
\docsecondrev{The} visual difference between the layout of $k\geq 100$ 
and the exact solution is \docsecondrev{negligible} because the number of pixel differences is small and the overview visualizations are similar.

Based on the performance of our method on the approximation quality and efficiency, we set $k=100$ in OoDAnalyzer. 
With this default setup, our method usually takes about $0.6$ seconds to lay out $2000$ data points and it is enough to fulfill the interactive visualization tasks as described in Sec.~\ref{subsec:interact}.


\subsection{Case Studies}\label{subsec:case}
We conducted two case studies to demonstrate the effectiveness of OoDAnalyzer for analyzing OoD samples in image classification. 
To verify the effectiveness of OoDAnalyzer, we expanded training data according to the identified OoD samples and then retrained the model.
An improvement of the model accuracy indicated that 
OoDAnalyzer successfully helped detect OoD samples and analyzed the underlying reason for their appearance.
In these case studies\gramsecondrev{,} we used 
pair analytics protocol~\cite{arias2011pair} \doc{in which} our collaborators drove the exploration and we navigated the system.  \looseness=-1

\subsubsection{OoD Analysis of the Animal Dataset.}
In this case study, we collaborated with experts $M_1$ and $M_2$ to effectively identify OoD samples in an animal classification task. 
Initially, the training dataset contains 5 categories, wolf, tiger, rabbit, dog, and cat,
\doc{which includes 1,562, 909, 1,253, 1,232, and 1,371 images, respectively}.
Correspondingly, the test dataset has 3,071, 1,895, 2,490, 2,509, and 2,317 images in each category.
After we identified the OoD samples in the initial datasets and improved the predictive performance, more samples and a new category,
leopard, was introduced, which also brought in more OoD samples. \looseness=-1

\myparagraph{OoD analysis in initial data (R1, R2, R3).} 
After loading the initial data into OoDAnalyzer, an overview of the model performance was provided.  
$M_1$ started the analysis with \docsecondrev{the} test data. 
He increased the cutoffs of the OoD score color bins to 0.6 and 0.8
by using the slider shown in Fig.~\ref{fig:teaser}F, 
so that the most significant OoD \changjian{samples} stood out. 
By looking at the overall distribution shown in Fig.~\ref{fig:teaser}, he found 4 areas of interest: region A, mixing cats and dogs; region B, misclassifying cartoon characters as rabbits; region C, confusing huskies with wolves; \doc{and} region D, mixing cats with tigers. 
$M_1$ decided to check these regions one by one as they all contained OoD \changjian{samples}.\looseness=-1

\begin{figure}[b]
  \centering
  \begin{overpic}[width = \linewidth]{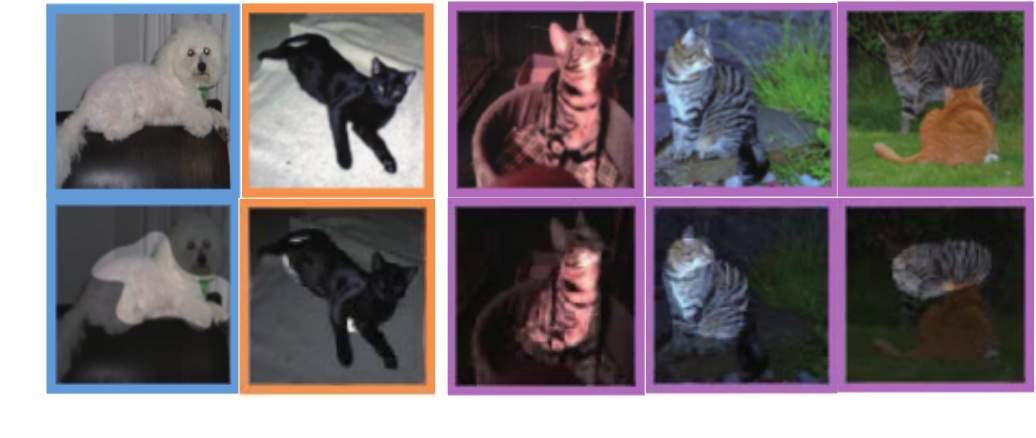}
  \put(18,-0.5){(a)}
  \put(69,-0.5){(b)}
  \put(1, 24){\begin{sideways} {\footnotesize OoD samples}\end{sideways}}
  \put(1, 4){\begin{sideways} {\footnotesize Saliency maps}\end{sideways}}
  \end{overpic}
    \caption{OoD samples in initial data \shixia{(top)} with their corresponding saliency maps \shixia{(bottom)}: (a) a white dog and a black cat misclassified as each other in region A; (b) three samples of cats with black stripes on their body misclassified as tigers in region D.}
    \label{fig:case1-saliency-map} 
\end{figure}

$M_1$ first zoomed into region A to understand why some \textbf{cats and dogs} were mixed together. 
He found that most misclassified images were light-colored dogs and dark-colored cats.
He examined the saliency maps and found that the effective parts learned by the model covered the whole body of the cats/dogs (Fig.~\ref{fig:case1-saliency-map}(a)), 
which indicated that the model might mistake colors as a criterion to distinguish cats and dogs.
To confirm the hypothesis, $M_1$ selected these two categories and switched to the juxtaposition mode to compare the distribution of \docsecondrev{the} training and test data.
Unsurprisingly, in \docsecondrev{the} training data, most dogs were \doc{dark-colored} and most cats were \doc{dark-colored},
while in test data, \docsecondrev{the} cats and dogs \doc{had} different colors.
Consequently, $M_1$ believed that a lack of diversity in color between cats and dogs was a major cause of these OoD samples.

\docsecondrev{He} continued the analysis with region B. Here a cluster of \textbf{cartoon images} of cats and dogs were misclassified as rabbits.
Again he examined the saliency maps and noticed that the ears of the cartoon characters were usually the key criteria for classification (Fig.~\ref{fig:teaser}(d)). 
\docsecondrev{To learn more about the} misclassification in the context of \docsecondrev{the} training samples, $M_1$ turned to the superposition mode.
There were no cartoon images in \docsecondrev{the} training data around the misclassified cartoon images, but many images of rabbits with prominent ears were presented,
which were similar to the exaggerated ears of cartoon images.
The model probably recognized these exaggerated ears as a symbol for rabbits, which led to the misclassification.
Thus, he indicated that a lack of cartoon images could account for such OoD samples. \looseness=-1

Next, $M_1$ moved to region C. This is at the center of the images predicted to be wolves, but many significant OoD \changjian{samples} also appeared.
Zooming into this region, $M_1$ saw that dogs were completely confused with wolves (Fig.~\ref{fig:teaser}(c)).
To figure out why the model predicted these dogs as wolves,
$M_1$ checked the nearest neighbors of these OoD \changjian{samples} and found many \textbf{huskies}, 
a kind of \gramsecondrev{dog} that looks like a wolf.
Given this finding, \changjian{he switched to \docsecondrev{the} training data (Fig.~\ref{fig:teaser}E)} to see if huskies were also placed close to the wolves.
However, he \docsecondrev{found very few} huskies in the training data
\docsecondrev{in either} the clusters of \cjsecondrev{the} dogs \docsecondrev{or} of \docsecondrev{the} wolves.
Therefore, $M_1$ concluded that a lack of huskies was the major reason for the appearance of these OoD samples.

Finally, $M_1$ noticed that some of the OoD \changjian{samples} were cats misclassified as tigers in region D.
To check this misclassification in detail, he reloaded the layout with only cats and tigers, and more OoD \changjian{samples} in these two categories were displayed.
There were misclassified images of \textbf{tabby cats with black stripes on their bodies}.
Their saliency maps showed that the effective parts learned by the model were these black stripes (Fig.~\ref{fig:case1-saliency-map}(b)), 
which \doc{are} actually a common pattern \docsecondrev{on} tigers.
Then $M_1$ used the juxtaposition mode to compare the cats in \docsecondrev{the} training data with those in test data.
He discovered that most cats in \docsecondrev{the} training data had no obvious patterns on their bodies.
As a result, a lack of tabby cats was identified as an underlying cause for such OoD samples.

To address the problems of the OoD samples mentioned above, 
we successively expanded the training data with 44 \doc{dark-colored} cats and 48 \doc{light-colored} dogs, 51 cartoon images, 
400 huskies \doc{and} 400 tabby cats.
The classification accuracy increased from 79.56\% to \textbf{84.01\%}, \textbf{85.58\%}, \textbf{88.23\%}, and \textbf{89.12\%}, respectively.

\begin{figure}[t]
 \centering
  \begin{overpic}[width = \linewidth]{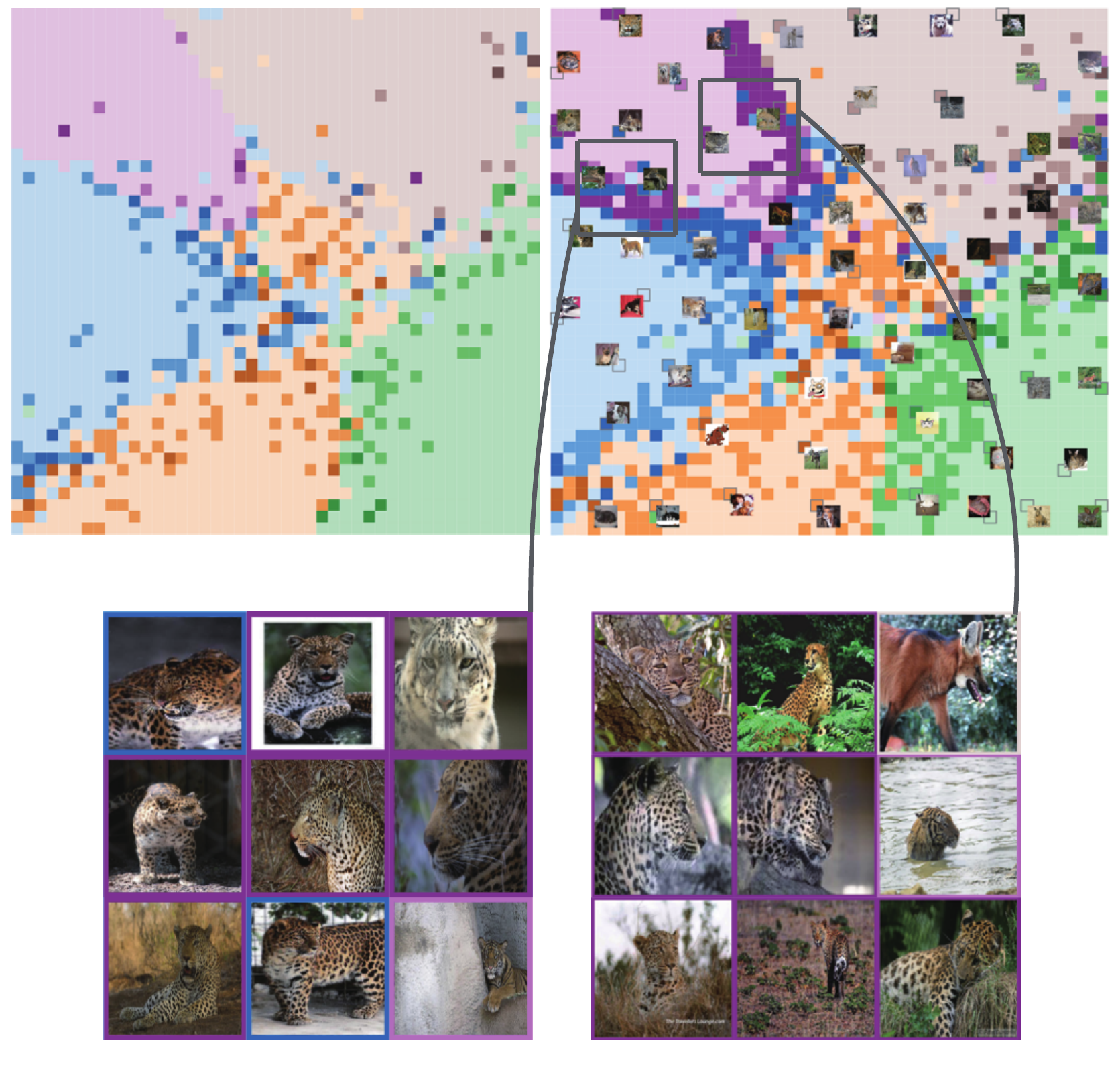}
  \put(24,43){(a)}
  \put(72,43){(b)}
  \put(24,-2){\large \contour[32]{white}{\textcolor{black}{A}}}
  \put(72,-2){\large \contour[32]{white}{\textcolor{black}{B}}}
  \end{overpic}
   \caption{Overview of test data: 
   (a) after extending the initial data according to the identified OoD samples; (b) more samples including leopards were added.}
    \label{fig:case1-retrained-overview} 
    \vspace{-3mm}
\end{figure}

After retraining the model, $M_1$ updated the layout and viewed it in our system.
As shown in Fig.~\ref{fig:case1-retrained-overview}(a), the overall color of the layout was much lighter than before with the same cutoffs of the color bins (Fig.~\ref{fig:teaser}(b)).
This change indicated that the number of OoD samples was reduced, which illustrated the usability of our system. \looseness=-1

\myparagraph{OoD analysis in incoming data (\textbf{R1}, \textbf{R2}).}
In real-world applications, 
new data flows in over time, \doc{which may} bring in new types of OoD samples, such as a new category. 
To illustrate how OoDAnalyzer handles new data, we \doc{introduced} more samples to the test dataset, as well as one more category, leopard.
With the expanded training dataset in the previous step,
the test dataset was enlarged to have 4,644, 2,794, 3,734, 3,847, and 3,394 images in each of the old categories and 1,000 images in the leopard category. \looseness=-1

\docsecondrev{When applying} our model 
\docsecondrev{to} the new test dataset, the accuracy degraded from 89.12\% to 84.21\%.
To diagnose the reason for the performance degradation, $M_2$ uploaded the datasets into OoDAnalyzer. 
The overview of \docsecondrev{the} test data showed two regions with high OoD scores (Fig.~\ref{fig:case1-retrained-overview}(b)). 
After zooming into these two regions, he discovered many leopard images hidden among the tigers (Fig.~\ref{fig:case1-retrained-overview}A, Fig.~\ref{fig:case1-retrained-overview}B).
Knowing that the training dataset had no leopards, he promptly identified leopard as a new category. 
Consequently, he added 400 leopard images to the training dataset and retrained the model.
The accuracy \docsecondrev{then} increased to \textbf{89.71\%}.\looseness=-1

\subsubsection{OoD analysis of Retinal Edema Dataset.}
This case study shows a real-world scenario \docsecondrev{in which} OoDAnalyzer \docsecondrev{is used to} identify OoD samples 
in a retinal edema dataset.

Over the past two years, our collaborators, two clinical doctors in ophthalmology ($D_1$ and $D_2$), have been applying machine learning techniques to OCT (optical coherence tomography) images to identify REA (retinal edema area) and classify images into two categories: REA and REA-free.
In their trials, the doctors often observed model performance degradation 
\docsecondrev{in}
the test dataset and they wanted to explore the reason.
Thus, we presented them with a visualization of their dataset, 
in which the training data contain\doc{ed} 8,960 OCT images and the test data contain\doc{ed} 1,920 OCT images. 
We used the UNet++ model from an award-winning implementation in AI Challenger 2018~\cite{wangshen2019}. \looseness=-1

\begin{figure}[t]
  \centering
  \begin{overpic}[width =\linewidth]{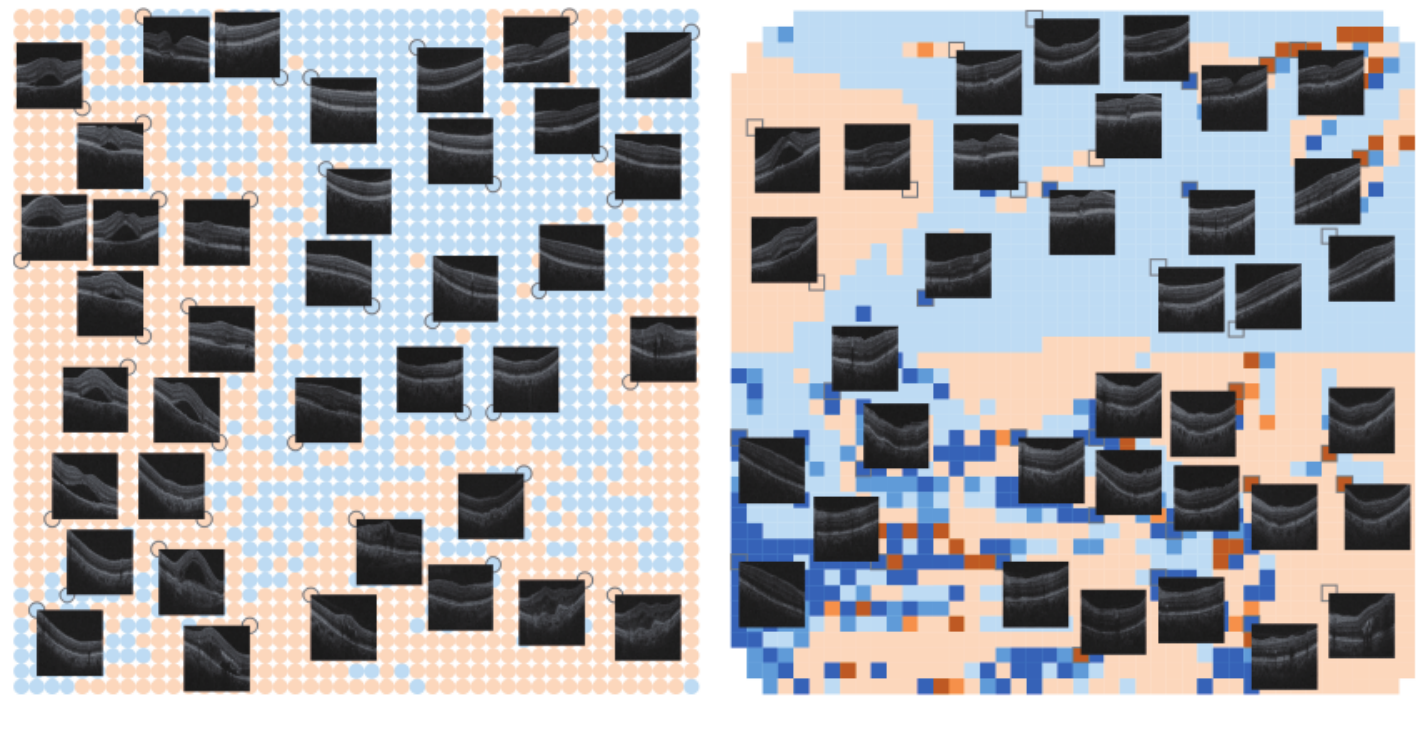}
  \put(19,-1){(a)}
  \put(74,-1){(b)}
  \put(29,35){\large \contour[32]{white}{\textcolor{black}{A}}}
  \put(1,21){\large \contour[32]{white}{\textcolor{black}{B}}}
  \put(27.2,4){\large \contour[32]{white}{\textcolor{black}{C}}}
  \end{overpic}
  \caption{The REA dataset: (a) training data; (b) test data.}
    \label{fig:case2-juxta}  
    \vspace{-3mm}
\end{figure}

\myparagraph{Overview (R1, R2).} 
Initially, a single view of the test dataset was presented,
where REA and REA-free categories were not separated well and many samples had high OoD scores (Fig.~\ref{fig:case2-juxta}(b)). To compare the distribution of the test dataset with that of the  training dataset, he switched to the juxtaposition mode.
Because it was not as jumbled as the test dataset, the two categories were much more clearly separated in the training dataset (Fig.~\ref{fig:case2-juxta}(a)).
Looking at some representative images,
he found that the blue area mostly consisted of REA-free images (Fig.~\ref{fig:case2-juxta}A), the left orange area contained NRD (neurosensory retinal detachment) images (Fig.~\ref{fig:case2-juxta}B),
and the bottom-right orange area had many CME (cystoid macular edema) images (Fig.~\ref{fig:case2-juxta}C).
Seeing such a clear pattern in the training dataset but not in the test dataset, $D_1$ decided to use the superposition mode to examine 
\docsecondrev{the}
OoD \changjian{samples} in context. \looseness=-1

\begin{figure}[b]
  \centering
  \includegraphics[width =0.95 \linewidth]{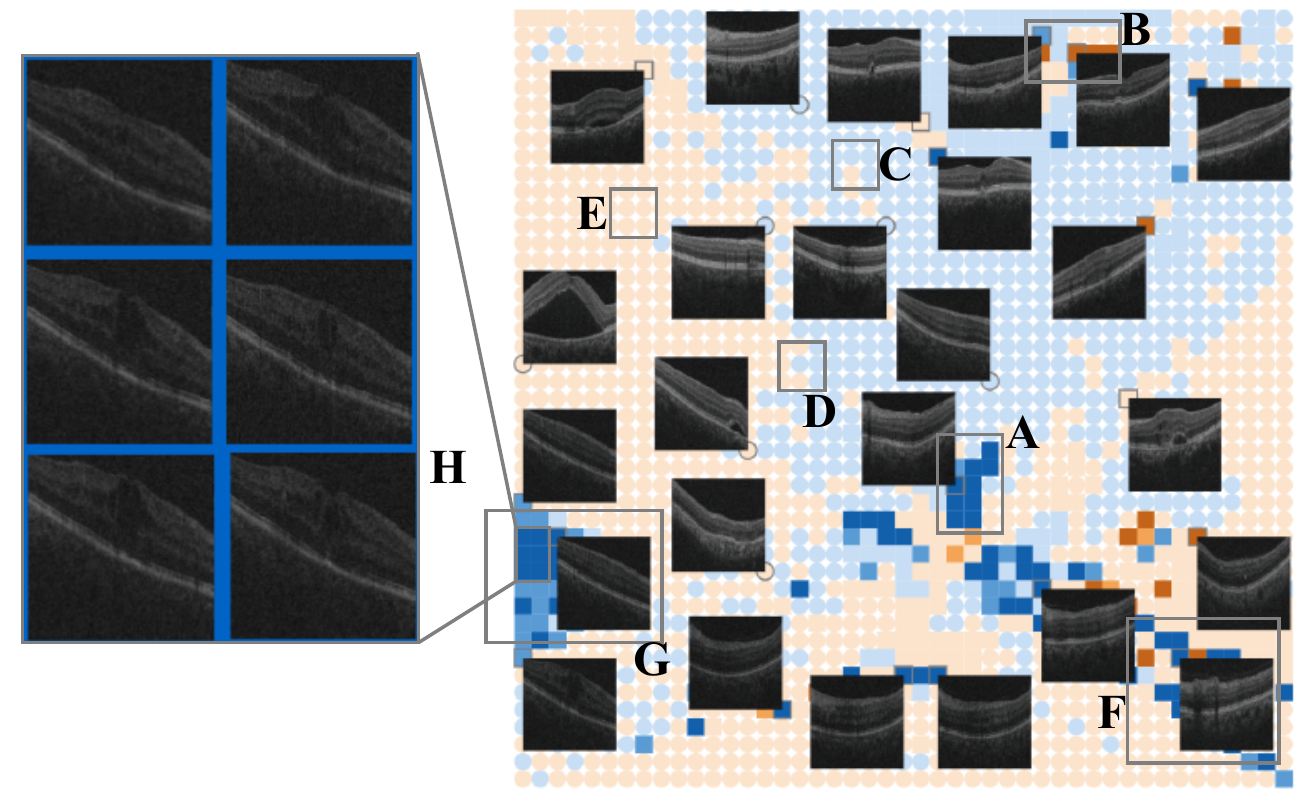}
    \vspace{-1mm}
    \caption{\changjian{The superposition mode of the REA dataset.}
    }
    \label{fig:sample-type} 
\end{figure}

\begin{figure}[t]
  \centering
  \begin{overpic}[width =\linewidth]{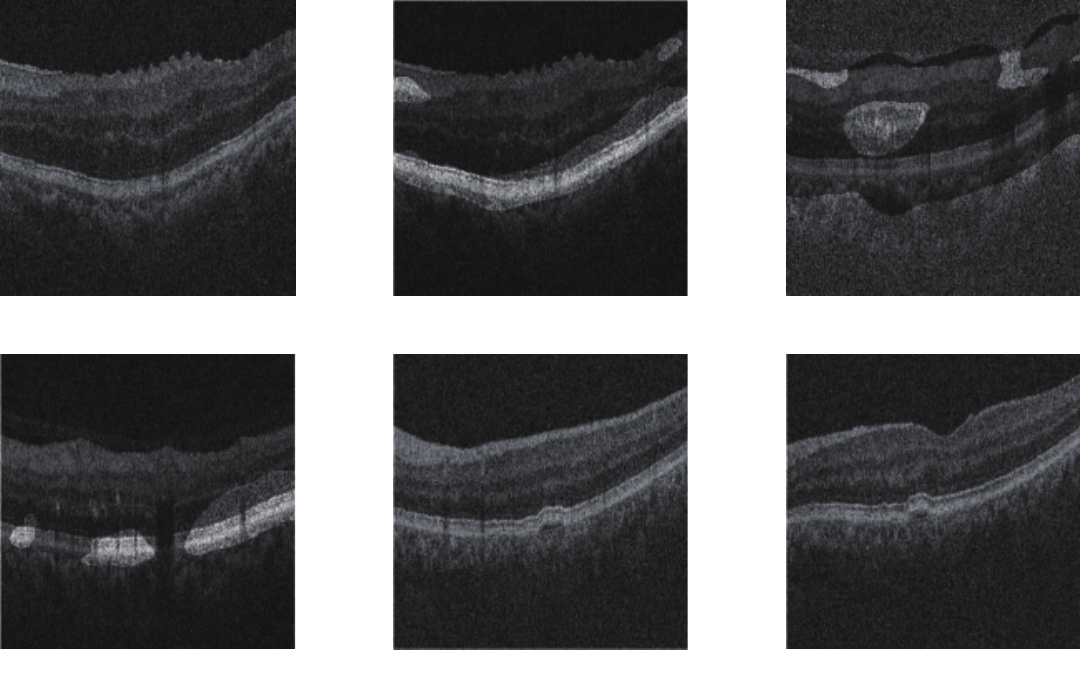}
  \put(12,33){(a)}
  \put(48,33){(b)}
  \put(85,33){(c)}
  \put(12,0){(d)}
  \put(48,0){(e)}
  \put(85,0){(f)}
  \end{overpic}
  \caption{Images and saliency maps: (a) a sample that had intraretinal cystoid cavities but no obvious retinal thickening; (b) the saliency map of (a); (c) the saliency map of a sample with obvious HRFs; (d) the saliency map of a sample with faint HRFs; (e) a misclassified sample with PED; (f) a correctly classified sample with PED. }
    \vspace{-4mm}
    \label{fig:case2-images} 
\end{figure}

\myparagraph{OoD \changjian{samples} with no obvious retinal thickening (R1, R2, R3).}
In the superposition mode, OoD samples \docsecondrev{were} gathered into smaller groups. 
There were four groups of samples with high OoD scores, marked as A, B, F, and G on Fig.~\ref{fig:sample-type}.
$D_1$ analyzed region A first.
The representative images in region A looked like REA-free at first glance.  
$D_1$ then zoomed in for a detailed analysis.
More images in the region were displayed and some images with no obvious retinal thickening were classified as REA-free (\eg, Fig.~\ref{fig:case2-images}(a)). 
However, they had intraretinal cystoid cavities, which is one of the symptoms of retinal edema.
$D_1$ examined these images carefully and confirmed \docsecondrev{they were} REA images.
To figure out the underlying reason, $D_1$ 
looked at their neighboring images in the training dataset and found no samples that had intraretinal cystoid cavities without obvious retinal thickening. 
Hence we hypothesized that a lack of REA samples with intraretinal cystoid cavities but an absence of retinal thickening could be a major cause for the OoD samples.
We checked a few saliency maps (\eg, Fig.~\ref{fig:case2-images}(b)) and found that the area of the intraretinal cystoid cavities was not learned by the model. 
This further verified 
the hypothesis. \looseness=-1

$D_1$ switched back to the overview in the superposition mode and quickly identified a similar image in region G.
It also had edema but no obvious retinal thickening.
Zooming into G, he found more such images in the test dataset (Fig.~\ref{fig:sample-type}H). 
So far, $D_1$ \docsecondrev{had} confirmed the above hypothesis. 

To improve the model performance,
$D_1$ collected 57 extra OCT images with edema that had intraretinal cystoid cavities but no obvious retinal thickening.
We then performed data augmentation on these images and got 640 images in total, which were added to the training data. 
After retraining the model, the accuracy 
was improved from 90.47\% to \textbf{93.02\%}.  

\myparagraph{OoD \changjian{samples} with hyper-reflective foci (R1, R2).}
Next, $D_2$ began to analyze the OoD \changjian{samples} in region F by zooming into the second level. 
The images showed that these OoD \changjian{samples} all had HRF (hyper-reflective foci), which is auxiliary evidence for a retinal edema diagnosis.
Some of these images were predicted correctly as REA, while others were not. 
By examining their saliency maps, $D_2$ found that the model learned the HRF characteristic and made correct predictions when the HRF was obvious (\eg, Fig.~\ref{fig:case2-images}(c)), but failed to learn faint HRFs (\eg, Fig.~\ref{fig:case2-images}(d)).
Using the superposition mode, $D_2$ explored the neighboring images in the training dataset and found that their HRFs were all obvious. 
Therefore, $D_2$ identified that the training dataset missed samples of REA with faint HRFs that appeared in the test dataset.

To overcome such a mismatch between training and test data,
$D_2$ added 512 images by augmenting 63 REA samples with faint HRF to the training data.
The model was retrained and the accuracy \doc{improved to} \textbf{94.53\%}. \looseness=-1

\myparagraph{Pseudo OoD samples caused by a labeling error (R1, R3).}
Finally, $D_2$ analyzed the OoD \changjian{samples} in region B. 
He found that some PED (pigment epithelial detachment) samples were incorrectly classified as REA-free (\eg, Fig.~\ref{fig:case2-images}(e)), and some PED samples were correctly classified as REA (\eg, Fig.~\ref{fig:case2-images}(f)).
These misclassified samples showed no common patterns.
To figure out the underlying reason, $D_2$ examined their nearest training samples and found 3 out of 16 images were mislabeled.
$D_2$ suspected that these misclassifications might have been caused by incorrect labels.
He then checked more training samples close to these OoD samples and found some of them were indeed mislabeled.
As a result, $D_2$ concluded that these samples were not OoD samples.
After spotting this labeling error, $D_1$ and $D_2$ thoroughly checked the training dataset and confirmed that there were no other labeling \doc{errors}. \looseness=-1


%% file: discussion.tex
\section{Expert Feedback}
\changjian{
After the case studies, we gathered feedback from our collaborators.} 
Overall, they were quite satisfied with OoDAnalyzer. \yafeng{They liked that the images were displayed and similar ones were close. 
This saved them a lot of time and they did not have to traverse all \docsecondrev{the} images to find the misclassified ones.
} 
\yafeng{They also said that}
OoDAnalyzer helped them understand the inner workings of the machine learning model from a data perspective~\cite{liu2018steering}. 
For example, $D_2$ commented, ``The intuitive interface 
\yafeng{tells}
more about machine learning techniques and how to use them effectively.
The machine learning model works like a baby. It does things in the way that it is taught by training data.''

They also gave us constructive feedback for improvements.
Our collaborators commented that it \docsecondrev{takes a while}
\yafeng{to train the}
OoD detection model. 
One machine learning expert suggested that we can accelerate the low-level feature extraction with distributed computing, as \yafeng{this} is the most time-consuming part.
In addition, two doctors expressed \doc{the need} to analyze text data with OoDAnalyzer. They commented, \cjsecondrev{``}This system has shown its effectiveness in identifying and analyzing OoD images. It would be more useful if OoDAnalyzer can support the visualization of text data because there are massive medical records in our hospital.\cjsecondrev{''} 

\changjian{
During the feedback session, one question raised by our collaborators was \doc{whether OoDAnalyzer could help them identify more types of samples apart from known unknowns and unknown unknowns.}
With a thorough analysis of the OoD detection results as well as a systematic visual exploration of the samples,
we have identified three other types of samples: \emph{reliable samples}, \emph{normal samples}, and \emph{boundary samples}.
The classification is based on prediction confidence (how confident the model is about its prediction) and the OoD score (the OoD degree of a sample).
Fig.~\ref{fig:sample-type} visually illustrates the major characteristics of these five types of samples. \looseness=-1
}

\vspace{0mm}
\begin{itemize}[leftmargin=*]\setlength\itemsep{0mm}
    \item[--] \textbf{Known unknowns} are OoD samples with low confidence predictions and high OoD scores. This is evidenced by the fact that they are near the decision boundaries in the grid visualization (
    Fig.~\ref{fig:sample-type}A).  
    \item[--] \textbf{Unknown unknowns} are OoD samples with high confidence predictions and high OoD scores, 
    which are blended in \docsecondrev{with} normal samples with high confidence and are hard to recognize (
    Fig.~\ref{fig:sample-type}B).  
    \item[--] \textbf{Reliable samples} have high confidence scores and low OoD scores. They are generally far away from the boundaries between classes (
    Fig.~\ref{fig:sample-type}E). 
    \item[--]\textbf{Normal samples} have low OoD scores and relatively high confidence predictions, but not as confident as reliable samples (
    Fig.~\ref{fig:sample-type}C). 
    \item[--]\textbf{Boundary samples} have low confidence predictions and low OoD scores. They are usually mixed with the known unknowns around the boundaries (
    Fig.~\ref{fig:sample-type}D). 
\end{itemize}

\shixia{The machine learning experts commented that these five types of samples helped them understand the characteristics of different OoD samples more clearly, which is probably useful \doc{for designing} a more generic OoD detection algorithm.} 

\section{Conclusion and Discussion}
\label{sec:discussion}
We have developed OoDAnalyzer, a \yuanjun{visual analysis} approach for identifying OoD samples and explaining their context.
OoDAnalyzer integrates an ensemble OoD detection method with a grid-based visualization to support the analysis of OoD samples from global patterns to local context. 
\yafeng{The developed OoD detection \changjian{method} employs \gramsecondrev{an} enriched feature set and algorithm set to achieve a better model capacity.}
To support real-time interactions and large dataset exploration, 
we implemented a $k$NN approximation for efficiently generating the grid layout in the grid-based visualization.
A set of quantitative experiments \doc{were conducted} to demonstrate the efficiency and effectiveness of our OoD detection algorithm and $k$NN approximation algorithm. 
In addition, we conducted two case studies to illustrate how OoDAnalyzer can be used to analyze OoD samples in context and how machine learning experts and domain practitioners can use this approach to identify OoD samples and overcome the problems they cause for a better performance. 

Other than OoD analysis, the algorithms proposed in this paper can be 
\yafeng{generalized}
to other cases. For example, the $k$NN-based algorithm can speed up the layout process in many application scenarios, such as image browsing~\cite{fried2015isomatch, quadrianto2010beyond}, graph drawing~\cite{yoghourdjian2015high, marriott2012memorability}, and biochemical network analysis~\cite{li2005grid, kojima2008fast}, where grid layouts are used to display images, grouped networks, and keywords. \looseness=-1

While the usefulness of OoDAnalyzer is demonstrated in the evaluation, it comes with several limitations, which may open up opportunities for future research.

\myparagraph{Parameter sensitivity}. 
Theoretically, in ensemble methods, the more models there are, the better the performance. 
However, in practice, we found that more models with inappropriate parameter settings often resulted in poor performance. 
This is because inappropriate parameter settings often lead to a plethora of low confidence models, which undoubtedly hurt the performance of the ensemble method~\cite{fern2003online}.  
As a result, a question arises as to how many sets of parameters are enough and how to select the parameters for each set.  
We believe this is a promising venue for future research. \looseness=-1

\myparagraph{Effectively learning $k$ for the $k$NN algorithm}.
In the $k$NN-based grid layout algorithm, $k$ is a key factor to balance efficiency and layout effectiveness.
In our implementation, $k$ is empirically set as 100. 
To improve the extensibility of the graph layout algorithm, it would be interesting to study how to automatically or semi-automatically search for the best $k$ value.

\myparagraph{Generalization}. 
In our current implementation, we take image data as an example to illustrate the basic idea. 
Our approach can be directly used to analyze video data if a representative image can be extracted from each video.
As for textual data, another widely used data type in many applications, the basic pipeline and OoD detection method can be directly utilized. 
The main obstacle to the generalization is the sample visualization.
As a result, how to effectively visualize text data, especially OoD samples and their context, is a challenge \doc{that} is worth studying in the future.

\myparagraph{Other factors affecting model performance.}
\cjsecondrev{
In this paper, we \docsecondrev{have improved} \xia{the} model performance from a data perspective, i.e., OoD samples. 
However, identifying the root cause of these OoD samples and adding more training samples \docsecondrev{does not} \xia{always} guarantee a better performance.
The feature and model used for classification also play important roles.
\xia{As a result, how to integrate our method with feature engineering and interactive model analysis} is an interesting venue for future work.
}

%% file: appendix.tex
\section*{Appendix A: Proof on the correctness of the greedy modification}
The proof includes two parts: the greedy modification can terminate, and the resulting graph satisfies the marriage theorem.

\vspace{1mm}
\noindent\emph{Proof.}
First, notice that every edge deletion operation reduces the degree of any vertex in $Y_>$, and every edge insertion operation increases the degree of any vertex in $Y_<$. When the degree of a vertex in $Y_>$ is reduced to $k$, it will be removed from $Y_>$. 
As the sizes of $Y_>$ and $Y_<$ are bounded and reduced by edge deletion or insertion respectively, the greedy algorithm always terminates.

Next, we prove that the modification yields $Y_< = \varnothing$ and $Y_> = \varnothing$ by contradiction.
We analyze the possibility of three cases when the greedy modification terminates.
\begin{itemize}[leftmargin=*]\setlength\itemsep{-1mm}
\item[--]\textbf{Case 1}: $Y_> \neq \varnothing, Y_< \neq \varnothing$. 
When the algorithm terminates under this case, it means that for any $\my \in Y_>$, all of its neighboring vertices are connected with every vertex of $Y_<$. 
Otherwise, there exists a vertex $\my \in Y_>$ and one of its neighboring vertices $\mx$ has non-empty $U_\mx$. In other words, we can find a vertex $\my^\prime \in Y_<$ which is not connected with $\mx$. According to the greedy algorithm, we can remove edge $\overline{\mx \my}$ and connect $\my^\prime$ to $\mx$. But this modification conflicts with the assumption of algorithm termination.

When all of the neighboring vertices of $\my \in Y_>$ are connected with every vertex of $Y_<$, the degree of any vertex in $Y_<$ is equal to $\md(\my) $ which is greater than $k$. This result violates the definition of $Y_<$ clearly.
\item[--]\textbf{Case 2}: $Y_> \neq \varnothing, Y_< = \varnothing$.  Note that the greedy modification does not change the sum of vertex degrees of $G_r$, which is $2kN$.  However, for Case 2, the vertex set of the graph is $(X_, Y_>, Y_=)$, and the sum of vertex degrees is $\sum_{\mx \in X_=}\deg(\mx) + \sum_{\my \in Y_>}\deg(\my) + \sum_{\my \in Y_=}\deg(\my)$, which is greater than $kN + k |Y_>| + k |Y_=| = 2 k N$, so that it contradicts the property of the greedy modification.
\item[--]\textbf{Case 3}: $Y_> = \varnothing, Y_< \neq \varnothing$.  The vertex set is $(X_, Y_<, Y_=)$, and the sum of vertex degrees is $\sum_{\mx \in X_=}\deg(\mx) + \sum_{\my \in Y_<}\deg(\my) + \sum_{\my \in Y_=}\deg(\my)$, which is less than $kN + k |Y_<| + k |Y_=| = 2 k N$ and contradicts the greedy modification as well.$\qed$
\end{itemize}

\vspace{12mm}
\section*{Appendix B: The effect of the number of logistic regression models}
The number of logistic regression models is an important hyper-parameter of our ensemble OoD detection method. 
In the presented appendix, we evaluate its effect on the OoD detection performance.

The previous study has shown that the regularization coefficient of logistic regression is generally chosen from an exponentially growing sequence~\cite{scikit-learn}. 
As a result, the regularization coefficient is chosen from \{$10^k$: $k = -5,\cdots,-1,0, 1,\cdots,5$\}.
To better cover the parameter space, we try to select the values with the largest overall distance among the regularization coefficients of the logistic regression algorithms in each ensemble method.
For example, when the classifier number in the ensemble method is 3, we set the regularization coefficient as \{$10^{-5}, 1, 10^5$\}. In particular, we set the regularization coefficient as 1 when the classifier number in the ensemble method is 1. We then evaluated the ensemble OoD detection method with the classifier number changing from 1 to 11.

\begin{figure}[h]
  \centering
  \begin{overpic}[width=0.9\linewidth]{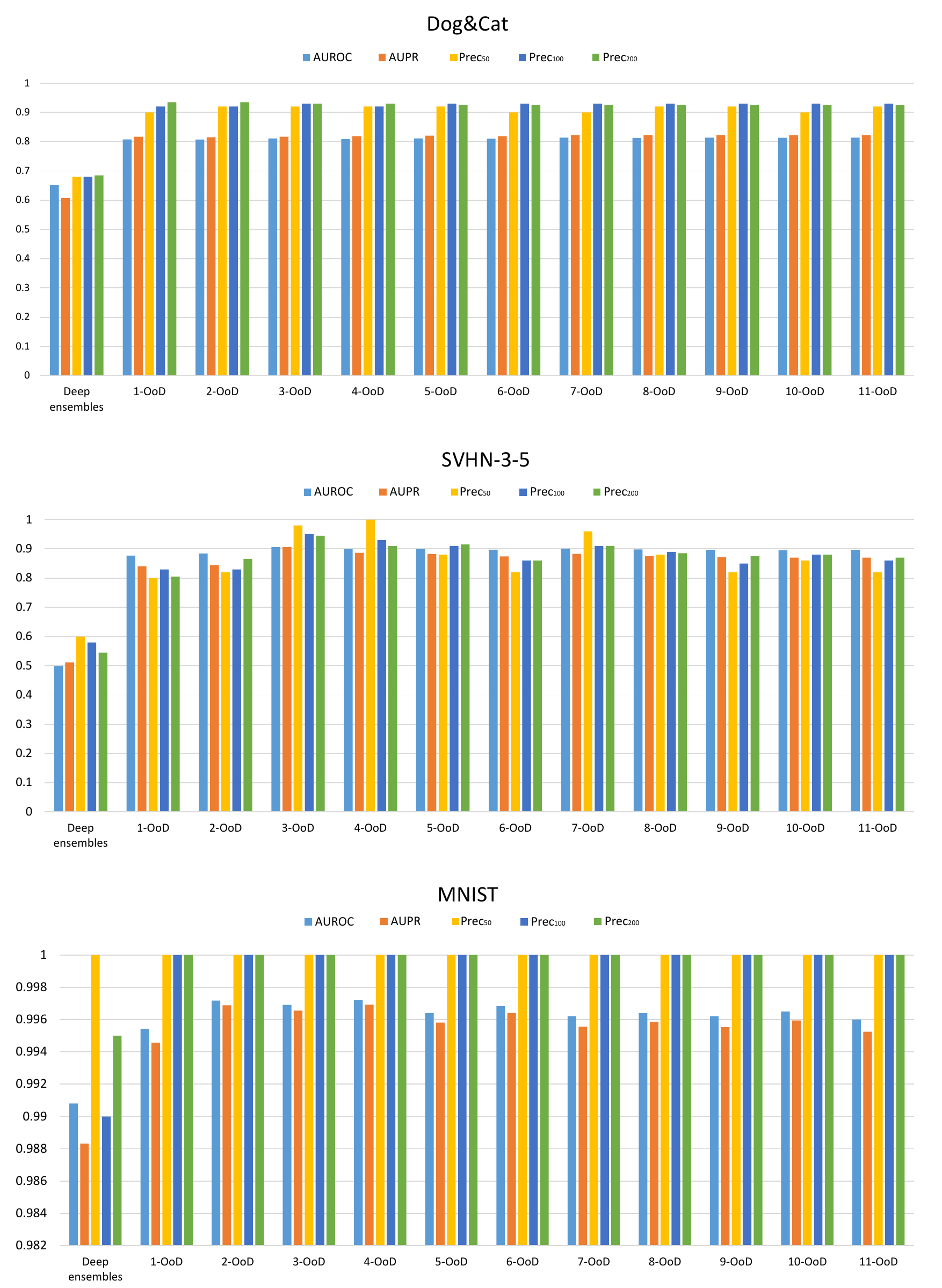}
  \end{overpic}
  \caption{Comparison of OoD detection performance among deep ensemble and our method with different numbers of logistic regression models.}\label{fig:varing-num}
\end{figure}

\emph{Result analysis.} Fig.~\ref{fig:varing-num} illustrates the OoD performance of our method with different numbers of logistic regression models.
Here $n$-OoD represents the ensemble method with $n$ models.
It can seen from Fig.~\ref{fig:varing-num} that even with only one model, our method is clearly better than the deep ensembles method. 
In particular, when the classifier number is greater than or equal to 3, the ensemble OoD detection method achieves an acceptable result. 
Thus, we set the classifier number as 3 in OoDAnalyzer.

%% file: OoDAnalyzer-TVCG.bbl
\begin{thebibliography}{10}
\providecommand{\url}[1]{#1}
\csname url@samestyle\endcsname
\providecommand{\newblock}{\relax}
\providecommand{\bibinfo}[2]{#2}
\providecommand{\BIBentrySTDinterwordspacing}{\spaceskip=0pt\relax}
\providecommand{\BIBentryALTinterwordstretchfactor}{4}
\providecommand{\BIBentryALTinterwordspacing}{\spaceskip=\fontdimen2\font plus
\BIBentryALTinterwordstretchfactor\fontdimen3\font minus
  \fontdimen4\font\relax}
\providecommand{\BIBforeignlanguage}[2]{{%
\expandafter\ifx\csname l@#1\endcsname\relax
\typeout{** WARNING: IEEEtran.bst: No hyphenation pattern has been}%
\typeout{** loaded for the language `#1'. Using the pattern for}%
\typeout{** the default language instead.}%
\else
\language=\csname l@#1\endcsname
\fi
#2}}
\providecommand{\BIBdecl}{\relax}
\BIBdecl

\bibitem{hendrycks2016baseline}
D.~Hendrycks and K.~Gimpel, ``A baseline for detecting misclassified and
  out-of-distribution examples in neural networks,'' in \emph{Proceedings of
  International Conference on Learning Representations}, 2017.

\bibitem{louizos2017multiplicative}
C.~Louizos and M.~Welling, ``Multiplicative normalizing flows for variational
  bayesian neural networks,'' in \emph{Proceedings of the International
  Conference on Machine Learning}, 2017, pp. 2218--2227.

\bibitem{guo2017calibration}
C.~Guo, G.~Pleiss, Y.~Sun, and K.~Q. Weinberger, ``On calibration of modern
  neural networks,'' in \emph{Proceedings of the International Conference on
  Machine Learning}, 2017, pp. 1321--1330.

\bibitem{lee2018simple}
K.~Lee, K.~Lee, H.~Lee, and J.~Shin, ``A simple unified framework for detecting
  out-of-distribution samples and adversarial attacks,'' in \emph{Proceedings
  of the Advances in Neural Information Processing Systems}, 2018, pp.
  7167--7177.

\bibitem{liang2017enhancing}
S.~Liang, Y.~Li, and R.~Srikant, ``Enhancing the reliability of
  out-of-distribution image detection in neural networks,'' in
  \emph{Proceedings of International Conference on Learning Representations},
  2018.

\bibitem{lakkaraju2017identifying}
H.~Lakkaraju, E.~Kamar, R.~Caruana, and E.~Horvitz, ``Identifying unknown
  unknowns in the open world: Representations and policies for guided
  exploration.'' in \emph{Proceedings of AAAI Conference on Artificial
  Intelligence}, 2017, pp. 2124--2132.

\bibitem{lakshminarayanan2017simple}
B.~Lakshminarayanan, A.~Pritzel, and C.~Blundell, ``Simple and scalable
  predictive uncertainty estimation using deep ensembles,'' in
  \emph{Proceedings of the Advances in Neural Information Processing Systems},
  2017, pp. 6402--6413.

\bibitem{lee2017training}
K.~Lee, H.~Lee, K.~Lee, and J.~Shin, ``Training confidence-calibrated
  classifiers for detecting out-of-distribution samples,'' in \emph{Proceedings
  of International Conference on Learning Representations}, 2018.

\bibitem{barnett1974outliers}
V.~Barnett and T.~Lewis, \emph{Outliers in statistical data}.\hskip 1em plus
  0.5em minus 0.4em\relax Hoboken, USA: Wiley, 1974.

\bibitem{liu2014survey}
S.~Liu, W.~Cui, Y.~Wu, and M.~Liu, ``A survey on information visualization:
  Recent advances and challenges,'' \emph{The Visual Computer}, vol.~30,
  no.~12, pp. 1373--1393, 2014.

\bibitem{Jiang2018}
L.~Jiang, S.~Liu, and C.~Chen, ``Recent research advances on interactive
  machine learning,'' \emph{Journal of Visualization}, vol.~22, no.~2, pp.
  401--417, 2019.

\bibitem{liu2018analyzing}
M.~Liu, J.~Shi, K.~Cao, J.~Zhu, and S.~Liu, ``Analyzing the training processes
  of deep generative models,'' \emph{IEEE Transactions on Visualization and
  Computer Graphics}, vol.~24, no.~1, pp. 77--87, 2018.

\bibitem{lu2016exploring}
Y.~Lu, M.~Steptoe, S.~Burke, H.~Wang, J.-Y. Tsai, H.~Davulcu, D.~Montgomery,
  S.~R. Corman, and R.~Maciejewski, ``Exploring evolving media discourse
  through event cueing,'' \emph{IEEE Transactions on Visualization and Computer
  Graphics}, vol.~22, no.~1, pp. 220--229, 2016.

\bibitem{wang2019a}
H.~Wang, Y.~Lu, S.~T. Shutters, M.~Steptoe, F.~Wang, S.~Landis, and
  R.~Maciejewski, ``A visual analytics framework for spatiotemporal trade
  network analysis,'' \emph{IEEE Transactions on Visualization and Computer
  Graphics}, vol.~25, no.~1, pp. 331--341, 2019.

\bibitem{xu2017vidx}
P.~Xu, H.~Mei, L.~Ren, and W.~Chen, ``{ViDX}: Visual diagnostics of assembly
  line performance in smart factories,'' \emph{IEEE Transactions on
  Visualization and Computer Graphics}, vol.~23, no.~1, pp. 291--300, 2017.

\bibitem{zhao2014fluxflow}
J.~Zhao, N.~Cao, Z.~Wen, Y.~Song, Y.-R. Lin, and C.~Collins, ``\# {FluxFlow}:
  Visual analysis of anomalous information spreading on social media,''
  \emph{IEEE Transactions on Visualization and Computer Graphics}, vol.~20,
  no.~12, pp. 1773--1782, 2014.

\bibitem{cao2016targetvue}
N.~Cao, C.~Shi, S.~Lin, J.~Lu, Y.-R. Lin, and C.-Y. Lin, ``{TargetVue}: Visual
  analysis of anomalous user behaviors in online communication systems,''
  \emph{IEEE Transactions on Visualization and Computer Graphics}, vol.~22,
  no.~1, pp. 280--289, 2016.

\bibitem{cao2018zglyph}
N.~Cao, Y.-R. Lin, D.~Gotz, and F.~Du, ``{Z-Glyph}: Visualizing outliers in
  multivariate data,'' \emph{Information Visualization}, vol.~17, no.~1, pp.
  22--40, 2018.

\bibitem{thom2012spatiotemporal}
D.~{Thom}, H.~{Bosch}, S.~{Koch}, M.~{Wörner}, and T.~{Ertl}, ``Spatiotemporal
  anomaly detection through visual analysis of geolocated twitter messages,''
  in \emph{Proceedings of the IEEE Pacific Visualization Symposium}, 2012, pp.
  41--48.

\bibitem{wilkinson2018visualizing}
L.~Wilkinson, ``Visualizing big data outliers through distributed
  aggregation,'' \emph{IEEE Transactions on Visualization and Computer
  Graphics}, vol.~24, no.~1, pp. 256--266, 2018.

\bibitem{ko2014analyzing}
S.~Ko, S.~Afzal, S.~Walton, Y.~Yang, J.~Chae, A.~Malik, Y.~Jang, M.~Chen, and
  D.~Ebert, ``Analyzing high-dimensional multivariate network links with
  integrated anomaly detection, highlighting and exploration,'' in
  \emph{Proceedings of the IEEE Conference on Visual Analytics Science and
  Technology}, 2014, pp. 83--92.

\bibitem{cao2018voila}
N.~Cao, C.~Lin, Q.~Zhu, Y.-R. Lin, X.~Teng, and X.~Wen, ``Voila: Visual anomaly
  detection and monitoring with streaming spatiotemporal data,'' \emph{IEEE
  Transactions on Visualization and Computer Graphics}, vol.~24, no.~1, pp.
  23--33, 2018.

\bibitem{xie2019visual}
C.~Xie, W.~Xu, and K.~Mueller, ``A visual analytics framework for the detection
  of anomalous call stack trees in high performance computing applications,''
  \emph{IEEE Transactions on Visualization and Computer Graphics}, vol.~25,
  no.~1, pp. 215--224, 2019.

\bibitem{liu2019interactive}
S.~Liu, C.~Chen, Y.~Lu, F.~Ouyang, and B.~Wang, ``An interactive method to
  improve crowdsourced annotations,'' \emph{IEEE Transactions on Visualization
  and Computer Graphics}, vol.~25, no.~1, pp. 235--245, 2019.

\bibitem{liu2018bridging}
S.~Liu, X.~Wang, C.~Collins, W.~Dou, F.~Ouyang, M.~El-Assady, L.~Jiang, and
  D.~Keim, ``Bridging text visualization and mining: A task-driven survey,''
  \emph{IEEE Transactions on Visualization and Computer Graphics}, vol.~25,
  no.~7, pp. 2482--2504, 2019.

\bibitem{Torralba2011unbaised}
A.~Torralba and A.~A. Efros, ``Unbiased look at dataset bias,'' in
  \emph{Proceedings of the IEEE Conference on Computer Vision and Pattern
  Recognition}, 2011, pp. 1521--1528.

\bibitem{liu2017towards}
M.~Liu, J.~Shi, Z.~Li, C.~Li, J.~Zhu, and S.~Liu, ``Towards better analysis of
  deep convolutional neural networks,'' \emph{IEEE Transactions on
  Visualization and Computer Graphics}, vol.~23, no.~1, pp. 91--100, 2017.

\bibitem{donahue2014decaf}
J.~Donahue, Y.~Jia, O.~Vinyals, J.~Hoffman, N.~Zhang, E.~Tzeng, and T.~Darrell,
  ``De{CAF}: A deep convolutional activation feature for generic visual
  recognition,'' in \emph{Proceedings of the International Conference on
  Machine Learning}, 2014, pp. 647--655.

\bibitem{Wang2019Visual}
Q.~{Wang}, J.~{Yuan}, S.~{Chen}, H.~{Su}, H.~{Qu}, and S.~{Liu}, ``Visual
  genealogy of deep neural networks,'' \emph{IEEE Transactions on Visualization
  and Computer Graphics (accepted)}, 2019.

\bibitem{gleicher2011visual}
M.~Gleicher, D.~Albers, R.~Walker, I.~Jusufi, C.~D. Hansen, and J.~C. Roberts,
  ``Visual comparison for information visualization,'' \emph{Information
  Visualization}, vol.~10, no.~4, pp. 289--309, 2011.

\bibitem{gleicher2018considerations}
M.~Gleicher, ``Considerations for visualizing comparison,'' \emph{IEEE
  Transactions on Visualization and Computer Graphics}, vol.~24, no.~1, pp.
  413--423, 2018.

\bibitem{hoeting1999bayesian}
J.~A. Hoeting, D.~Madigan, A.~E. Raftery, and C.~T. Volinsky, ``Bayesian model
  averaging: a tutorial,'' \emph{Statistical science}, vol.~14, no.~4, pp.
  382--401, 1999.

\bibitem{goodfellow2016deep}
I.~Goodfellow, Y.~Bengio, and A.~Courville, \emph{Deep learning}.\hskip 1em
  plus 0.5em minus 0.4em\relax Massachusetts, USA: MIT press, 2016.

\bibitem{lowe2004distinctive}
D.~G. Lowe, ``Distinctive image features from scale-invariant keypoints,''
  \emph{International Journal of Computer Vision}, vol.~60, no.~2, pp. 91--110,
  2004.

\bibitem{rublee2011orb}
E.~Rublee, V.~Rabaud, K.~Konolige, and G.~Bradski, ``{ORB}: An efficient
  alternative to {SIFT} or {SURF},'' in \emph{Proceedings of the International
  Conference on Computer Vision}, 2011, pp. 2564--2571.

\bibitem{calonder2010brief}
M.~Calonder, V.~Lepetit, C.~Strecha, and P.~Fua, ``{BRIEF}: Binary robust
  independent elementary features,'' in \emph{Proceedings of the European
  Conference on Computer Vision}, 2010, pp. 778--792.

\bibitem{tsne}
L.~van~der Maaten and G.~Hinton, ``Visualizing data using t-{SNE},''
  \emph{Journal of Machine Learning Research}, vol.~9, no. Nov, pp. 2579--2605,
  2008.

\bibitem{schoeffmann2011similarity}
K.~Schoeffmann and D.~Ahlstrom, ``Similarity-based visualization for image
  browsing revisited,'' in \emph{Proceedings of the IEEE International
  Symposium on Multimedia}, 2011, pp. 422--427.

\bibitem{fried2015isomatch}
O.~Fried, S.~DiVerdi, M.~Halber, E.~Sizikova, and A.~Finkelstein, ``{IsoMatch}:
  Creating informative grid layouts,'' \emph{Computer Graphics Forum}, vol.~34,
  no.~2, pp. 155--166, 2015.

\bibitem{yoghourdjian2015high}
V.~Yoghourdjian, T.~Dwyer, G.~Gange, S.~Kieffer, K.~Klein, and K.~Marriott,
  ``High-quality ultra-compact grid layout of grouped networks,'' \emph{IEEE
  Transactions on Visualization and Computer Graphics}, vol.~22, no.~1, pp.
  339--348, 2015.

\bibitem{marriott2012memorability}
K.~Marriott, H.~Purchase, M.~Wybrow, and C.~Goncu, ``Memorability of visual
  features in network diagrams,'' \emph{IEEE Transactions on Visualization and
  Computer Graphics}, vol.~18, no.~12, pp. 2477--2485, 2012.

\bibitem{li2005grid}
W.~Li and H.~Kurata, ``A grid layout algorithm for automatic drawing of
  biochemical networks,'' \emph{Bioinformatics}, vol.~21, no.~9, pp.
  2036--2042, 2005.

\bibitem{kojima2008fast}
K.~Kojima, M.~Nagasaki, and S.~Miyano, ``Fast grid layout algorithm for
  biological networks with sweep calculation,'' \emph{Bioinformatics}, vol.~24,
  no.~12, pp. 1433--1441, 2008.

\bibitem{selvaraju2017grad}
R.~R. Selvaraju, M.~Cogswell, A.~Das, R.~Vedantam, D.~Parikh, and D.~Batra,
  ``{Grad-CAM}: Visual explanations from deep networks via gradient-based
  localization,'' in \emph{Proceedings of the International Conference on
  Computer Vision}, 2017, pp. 618--626.

\bibitem{Markovtsev2017}
V.~Markovtsev, ``{Jonker-Volgenant} algorithm + {t-SNE} = {Super} powers,''
  https://blog.sourced.tech/post/lapjv/, 2017, {L}ast accessed 2019-03-30.

\bibitem{isomap}
J.~B. Tenenbaum, V.~de~Silva, and J.~C. Langford, ``A global geometric
  framework for nonlinear dimensionality reduction,'' \emph{Science}, vol. 290,
  no. 5500, pp. 2319--2323, 2000.

\bibitem{jonker1987shortest}
R.~Jonker and A.~Volgenant, ``A shortest augmenting path algorithm for dense
  and sparse linear assignment problems,'' \emph{Computing}, vol.~38, no.~4,
  pp. 325--340, 1987.

\bibitem{munkres1957algorithms}
J.~Munkres, ``Algorithms for the assignment and transportation problems,''
  \emph{Journal of the Society for Industrial and Applied Mathematics}, vol.~5,
  no.~1, pp. 32--38, 1957.

\bibitem{dell2000algorithms}
M.~Dell'Amico and P.~Toth, ``Algorithms and codes for dense assignment
  problems: the state of the art,'' \emph{Discrete Applied Mathematics}, vol.
  100, no. 1-2, pp. 17--48, 2000.

\bibitem{bondy1976graph}
J.~A. Bondy and U.~S.~R. Murty, \emph{Graph theory with applications}.\hskip
  1em plus 0.5em minus 0.4em\relax London, UK: Macmillan, 1976.

\bibitem{leiserson2001introduction}
C.~E. Leiserson, R.~L. Rivest, T.~H. Cormen, and C.~Stein, \emph{Introduction
  to algorithms}.\hskip 1em plus 0.5em minus 0.4em\relax Massachusetts, USA:
  MIT Press, 2001.

\bibitem{xiang2019interactive}
S.~Xiang, X.~Ye, J.~Xia, J.~Wu, Y.~Chen, and S.~Liu, ``Interactive correction
  of mislabeled training data,'' in \emph{Proceedings of the IEEE Conference on
  Visual Analytics Science and Technology (accepted)}, 2019.

\bibitem{kaggle}
``Kaggle-dogs-vs-cats,'' https://www.kaggle.com/c/dogs-vs-cats, 2019, {L}ast
  accessed 2019-03-30.

\bibitem{netzer2011reading}
Y.~Netzer, T.~Wang, A.~Coates, A.~Bissacco, B.~Wu, and A.~Y. Ng, ``Reading
  digits in natural images with unsupervised feature learning,'' in \emph{NIPS
  Workshop on Deep Learning and Unsupervised Feature Learning}, 2011, pp. 1--9.

\bibitem{lecun1998gradient}
Y.~LeCun, L.~Bottou, Y.~Bengio, and P.~Haffner, ``Gradient-based learning
  applied to document recognition,'' \emph{Proceedings of the IEEE}, vol.~86,
  no.~11, pp. 2278--2324, 1998.

\bibitem{notmnist}
Y.~Bulatov, ``Notmnist dataset,''
  \url{https://yaroslavvb.blogspot.com/2011/09/notmnist-dataset.html}, 2011,
  {L}ast accessed 2019-03-30.

\bibitem{zhang2019learning}
L.~Zhang and S.~Rusinkiewicz, ``Learning local descriptors with a cdf-based
  dynamic soft margin,'' in \emph{Proceedings of the IEEE International
  Conference on Computer Vision}, 2019, pp. 2969--2978.

\bibitem{scikit-learn}
F.~Pedregosa, G.~Varoquaux, A.~Gramfort, V.~Michel, B.~Thirion, O.~Grisel,
  M.~Blondel, P.~Prettenhofer, R.~Weiss, V.~Dubourg, J.~Vanderplas, A.~Passos,
  D.~Cournapeau, M.~Brucher, M.~Perrot, and E.~Duchesnay, ``Scikit-learn:
  Machine learning in {P}ython,'' \emph{Journal of Machine Learning Research},
  vol.~12, pp. 2825--2830, 2011.

\bibitem{retinal}
``Ai-challenger-retinal-edema-segmentation,''
  \url{https://challenger.ai/dataset/fld2018}, 2018, {L}ast accessed
  2019-03-30.

\bibitem{russakovsky2015imagenet}
O.~Russakovsky, J.~Deng, H.~Su, J.~Krause, S.~Satheesh, S.~Ma, Z.~Huang,
  A.~Karpathy, A.~Khosla, M.~Bernstein, A.~C. Berg, and L.~Fei-Fei, ``{ImageNet
  Large Scale Visual Recognition Challenge},'' \emph{International Journal of
  Computer Vision}, vol. 115, no.~3, pp. 211--252, 2015.

\bibitem{krizhevsky2009learning}
A.~Krizhevsky and G.~Hinton, ``Learning multiple layers of features from tiny
  images,'' University of Toronto, Tech. Rep., 2009.

\bibitem{lapjv}
``Linear assignment problem solver using {Jonker-Volgenant} algorithm,''
  \url{https://github.com/src-d/lapjv}, 2017, {L}ast accessed 2019-03-30.

\bibitem{arias2011pair}
R.~Arias-Hernandez, L.~T. Kaastra, T.~M. Green, and B.~Fisher, ``Pair
  analytics: Capturing reasoning processes in collaborative visual analytics,''
  in \emph{Proceedings of the Hawaii International Conference on System
  Sciences}, 2011, pp. 1--10.

\bibitem{wangshen2019}
S.~Wang and Y.~Wang, ``Deepseg: Ai-challenger-retinal-edema-segmentation,''
  \url{https://github.com/ShawnBIT/AI-Challenger-Retinal-Edema-Segmentation},
  2018, {L}ast accessed 2019-03-30.

\bibitem{liu2018steering}
S.~Liu, G.~Andrienko, Y.~Wu, N.~Cao, L.~Jiang, C.~Shi, Y.-S. Wang, and S.~Hong,
  ``Steering data quality with visual analytics: The complexity challenge,''
  \emph{Visual Informatics}, vol.~2, no.~4, pp. 191--197, 2018.

\bibitem{quadrianto2010beyond}
N.~Quadrianto, K.~Kersting, T.~Tuytelaars, and W.~L. Buntine, ``{Beyond
  2D-grids}: A dependence maximization view on image browsing,'' in
  \emph{Proceedings of the International Conference on Multimedia Information
  Retrieval}, 2010, pp. 339--348.

\bibitem{fern2003online}
A.~Fern and R.~Givan, ``Online ensemble learning: An empirical study,''
  \emph{Machine Learning}, vol.~53, no. 1-2, pp. 71--109, 2003.

\end{thebibliography}
